\documentclass[10pt,aps,prx,showpacs,superscriptaddress,reprint,citeautoscript]{revtex4-1}

\usepackage{amssymb, amsmath,amsfonts}    
\usepackage{pifont}
\usepackage{amsmath}
\usepackage{color}      
\usepackage{graphicx}	
\usepackage{subfigure}
\usepackage{float}
\usepackage{amsmath}
\usepackage{comment}
\DeclareGraphicsExtensions{.eps,.png,.pdf,.jpg}

\def\eeq{\relax}
\def\beq#1#2\eeq{\begin{equation}\label{#1}#2\end{equation}}
\def\bal#1#2\eal{\begin{align}\label{#1}#2\end{align}}
\def\bse#1#2\ese{\begin{subequations}\label{#1}#2\end{subequations}}
\def\ba{\begin{aligned}}   \def\ea{\end{aligned}}

\def\dd{\operatorname{d}} 
\def\ii{\operatorname{i}}

 \newcommand{\rev}[1] {{#1}}

\begin{document}

\title{ Analytical solutions for single and multiple scattering \\ from  rib-stiffened plates in water}

\author{Hesam Bakhtiary Yekta and Andrew N. Norris}
\affiliation{Department of Mechanical \& Aerospace Engineering, Rutgers University, Piscataway, NJ 08854, USA}
\date{\today}

\begin{abstract}

The  interaction of an acoustic plane wave with a pair of plates connected by periodically spaced stiffeners in water is considered.  The rib-stiffened structure is called a ``flex-layer" because its low frequency response is dominated by  bending stiffness.   The quasi-static behavior  is equivalent   a homogeneous layer of compressible  fluid, which we identify as air for the purposes of comparison. In this way an  air layer is acoustically the same as a pair of  thin elastic plates connected by a periodic spacing of ribs.  
At  discrete higher  frequencies the flex-layer exhibits perfect acoustic transmission, the cause of  which is identified as fluid-loaded plate waves propagating back and forth between the ribs.  Both the low and finite frequency behavior of the flex-layer are fully explained by closed-form solutions for  reflection and transmission.   The analytical model is extended to two flex-layers in series, introducing new low and high frequency phenomena that are explained in terms of simple lumped parameter models. 

\end{abstract}

\maketitle

\section{Introduction} \label{sec1}   

\rev{Interaction of sound with rib-reinforced plates is a long-standing topic of interest related to  acoustical properties of walls, floors,  partitions,  sound absorbing panels and other structural elements.  Many different model configurations have been considered,  including:  
infinite plates with a single rib \cite{Lin1977a,Woolley1980a,Skelton1990a,Andronov2002}, multiple ribs \cite{Woolley1980},  periodic ribs \cite{Stepanishen1978,karali1994,Cray1994,xin2011transmission} or resonators \cite{Skelton2018}; finite ribbed plates   \cite{Keltie1993,TranVanNhieu2017,Wrona2020,Ye2022}, and fully three-dimensional models of a rib-stiffened plates  \cite{Hull2010, Remillieux2012, Xin2015}.  The objective here is an analytical solution for the     
transmission and reflection of underwater sound from two infinite thin plates in parallel with periodic rib stiffening.  Our interest in this structure was motivated by the observation that it behaves at low frequency as an equivalent spring-like layer.   This behavior contrasts  with the low frequency mass law observed for panels in air \cite{lin1977sound,wang2005sound}.  For this reason we  call the structure a ``flex-layer" because of its low frequency stiffness-dominated behavior. }

\rev{Our focus is the acoustic properties of a pair of infinitely long fluid-loaded plates connected by periodically spaced rib-stiffeners situated in an infinite acoustic fluid (water).   We develop closed-form solutions of the   reflection and transmission  for the pair of plates and also for a system comprising two pairs of plates separated by a water gap.  The solution for the latter $n=2$ system is found by considering the single scattering problem for an infinite set of evanescent plane waves that are related to the incident plane wave by the Bragg condition.  We obtain infinite sized  reflection and transmission matrices that can be combined to obtain the response from any number $n\ge 2$ of plate pairs.  
The fundamental solution method takes advantage of the infinite periodicity,  enabling closed-form solution through the use of the Poisson Summation formula, which has been used to advantage previously  in similar problems for vibration \cite{Rumerman1975}, sound radiation \cite{Evseev,mace1980sound,Takahashi1983} and reflection and transmission \cite{lin1977sound,Stepanishen1978,Skelton1990}.  }

The paper proceeds as follows.  The flex-layer model is introduced in Section \ref{sec2} and is shown to be acoustically equivalent at low frequency  to a layer of air in water.  A full frequency analytical solution for the acoustical response of the flex-layer is developed in Section \ref{sec3}.  \rev{The  solution as a sum of symmetric and antisymmetric partial solutions is, to our knowledge, novel, and leads to  expressions for equivalent plate impedances.}     The analytical model is compared with full scale simulation and its transmission properties are discussed in Section \ref{sec4}.
Transmission through a pair of flex-layers separated by water is considered in Section \ref{sec5} using a new type of  analytical solution.  Conclusions and future directions are presented in Section \ref{sec6}.

\section{Theory: Low frequency flex-layer} \label{sec2}   
The rib-stiffened structure,  of which a section is shown in Fig.\ \ref{fig1}(a), comprises  parallel plates a distance $l_r$ apart,  and periodically spaced ribs separated by $2b$.  
We are interested in how a plane wave incident from  the semi-infinite water region on the right is reflected and transmitted.  Before presenting the full solution in  Section \ref{sec3} 
we first give a simple solution for the scattering problem in the low frequency or ``quasi-static" frequency range.  

In the  quasi-static  regime the frequency is low enough that the dominant deformation mechanism of the plates is bending due to an imposed effectively static pressure, with the entrained air having negligible influence.  Hence the name ``flex-layer" for the rib-stiffened structure.   The deformation  between two ribs can be modeled as a clamped-clamped plate,  as in Fig.\ \ref{fig2}. The plate of thickness $h$ is under pressure $p_0$ on one side and subject to rigid line constraints
in the $x$-direction spaced a distance $2b$ apart (for the purposes of the quasi-static model the effect of a rib stiffener is assumed to be  approximated by a rigid constraint). The plate displacement  $w(y)$   satisfies the Euler-Bernoulli  plate equation (the static version of Kirchhoff plate theory) 
\beq{DWp0}
D  w^{\prime \prime \prime \prime}(y) = p_0,\quad\text{ $-b \le y \le b$}
\eeq
 where $D = {{E_p}I}/{(1-\nu^2)}$,  $I = {h^3}/{12}$, $E_p$ is the plate Young's modulus and $\nu$ the Poisson's ratio.
 \begin{widetext}      \begin{minipage}{\linewidth}    
  \begin{figure}[H]
    \centering
    \subfigure[\ Rib-stiffened plates]{\includegraphics[width=0.3\textwidth]{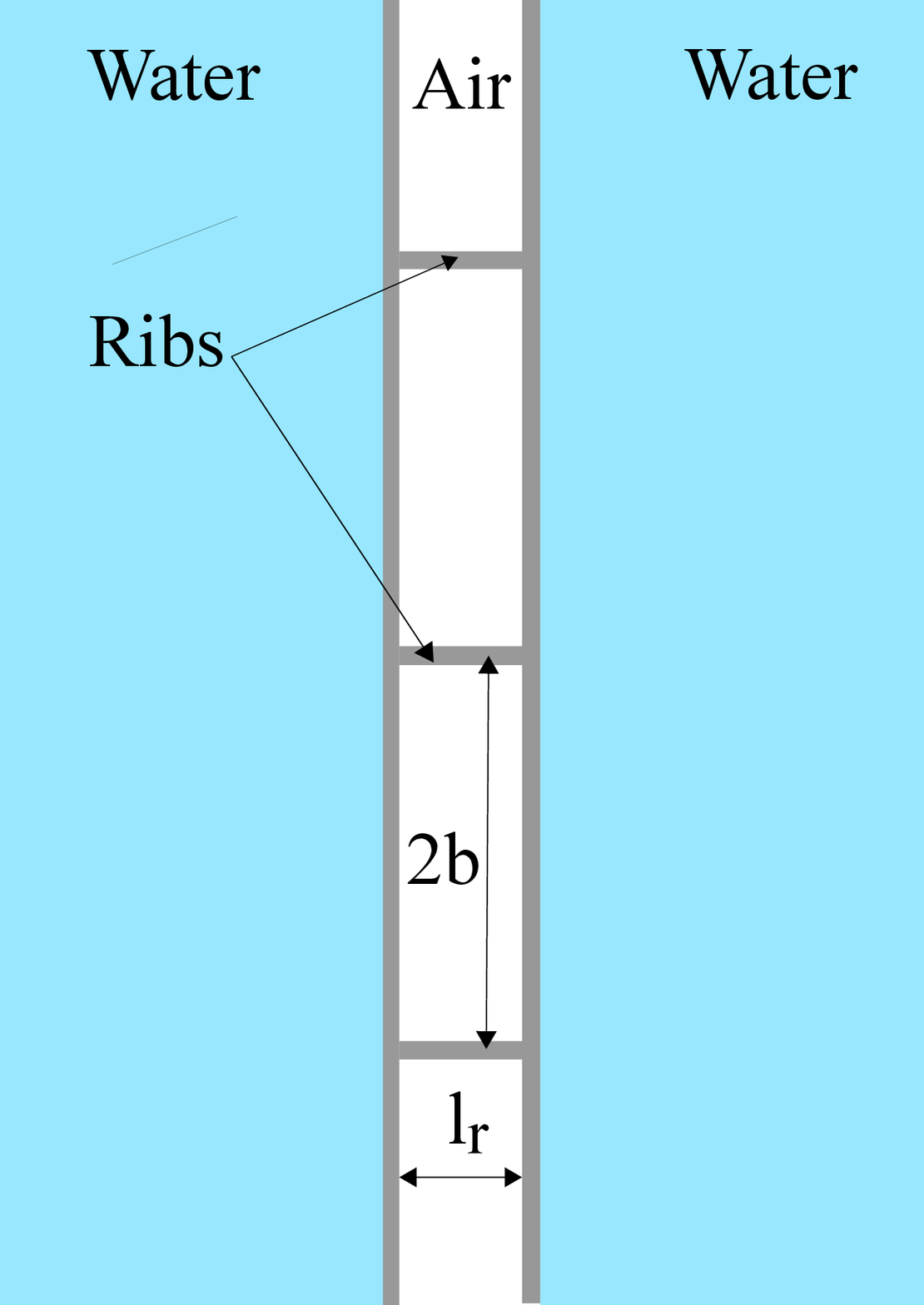}}
    \hspace{0.1\textwidth}
    \subfigure[\ Equivalent air gap at low frequency]{\includegraphics[width=0.3\textwidth]{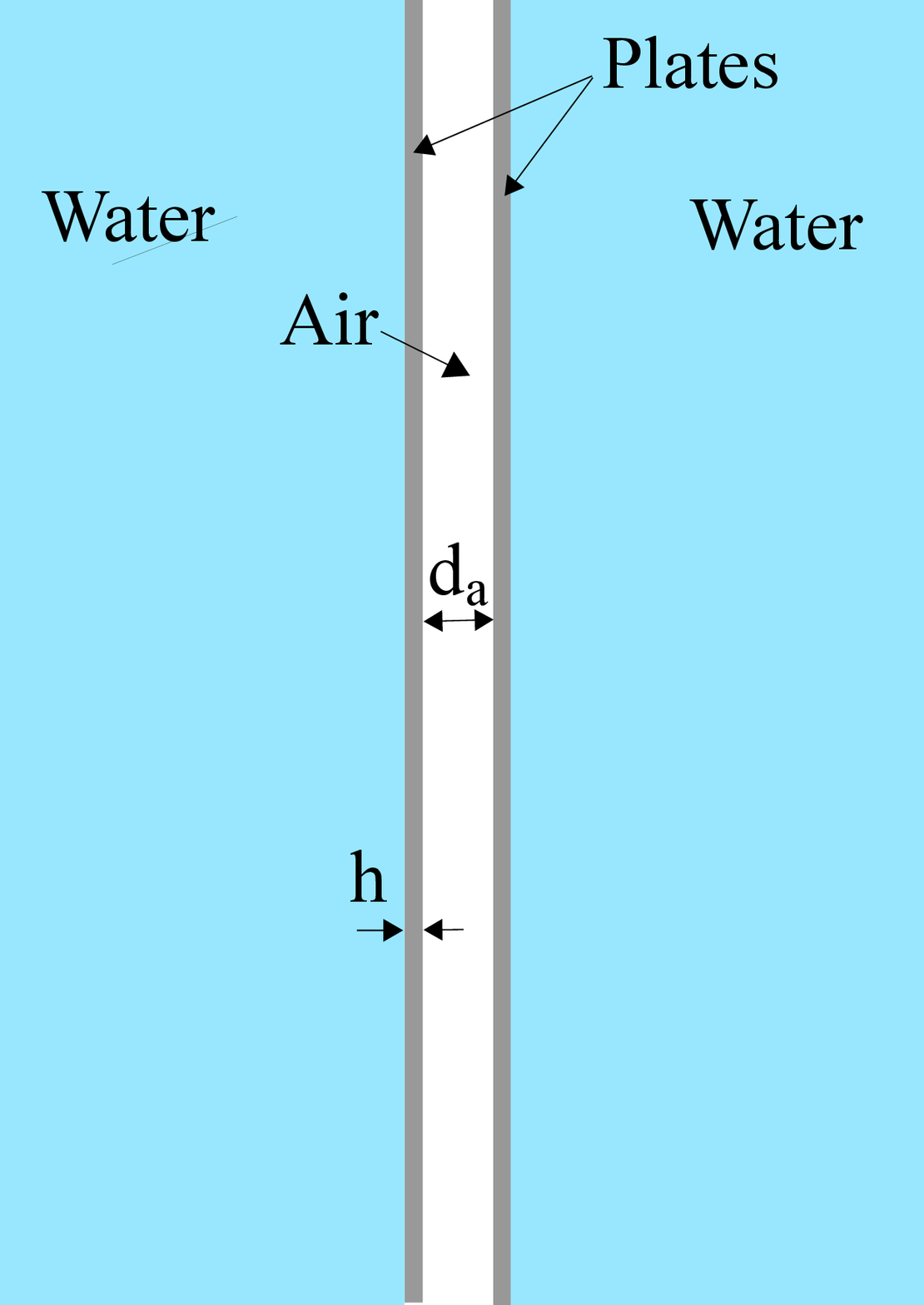} }
    \caption{ (a)  Flex-layer model of two plates with connecting ribs.  (b) Equivalent quasi-static air gap, including two thin elastic plates acting to maintain the air-water separation.  }
    \label{fig1}
\end{figure} 
    \end{minipage} \end{widetext}                     

\begin{figure}[H]
    \centering
    \includegraphics[width=0.35\textwidth]{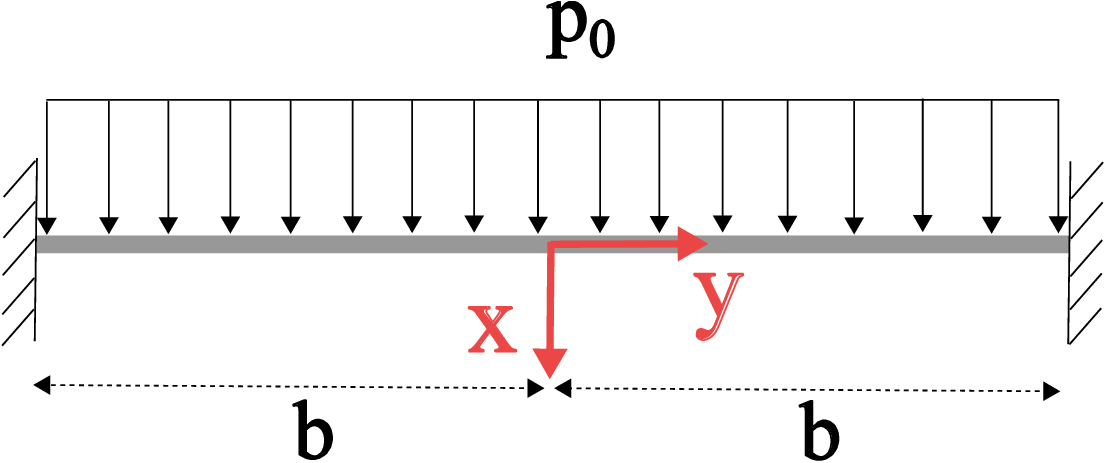}
    \caption{Clamped-clamped plate under a quasi-static pressure $p_0$. }
    \label{fig2}
\end{figure}
Solving Eq.\  \eqref{DWp0} with the  boundary conditions $w(\pm b) = 0$,  $w^{\prime }(\pm b) = 0$, yields  $w(y) = {p_0}\big(y^2 - b^2 \big)^2 /(24D) $. 
 Since there are two such plates with  rigid constraints connecting them, and
subject to pressure $p_0$ on both sides, the change in internal volume  over an area $A$ of the plates comprising a length in the $z$-direction and the span $-b \le y \le b$  is $2A w_{av}$, where $w_{av}$ is the average displacement.   The average change in distance separating the plates is therefore
$\Delta L = 2 w_{av}$, or 
\begin{equation}
\Delta L = \frac{2}{45} \frac{b^4}{D}p_0  .
\label{DVofFlex}
\end{equation}
A slightly larger value  is found for $\Delta L $ if the Timoshenko (or equivalently, Mindlin) plate theory is used instead of the Euler-Bernoulli theory.  The relative difference is, however, negligible;  see Appendix \ref{appa}.

The same $\Delta L$ could be obtained with a uniform ``spring-layer".   In order to make this  realistic, consider a thin layer of air separating two half-spaces of water, as shown in  Fig.\ \ref{fig1}(b).   The air gap acts at low frequencies as an effective linear spring, with stiffness proportional to the air bulk modulus, $K_a$, and inversely proportional to the gap width, $d_a$.  The additional rigid plates in Fig.\ \ref{fig1}(b) are introduced as separators between the air and water and do not change the main effect of the air gap  as an effective spring.   Low frequency here means that the acoustic wavelength in the air gap is much greater than the gap width $d_a$. This translates to frequencies $f \ll {c_a}/{d_a}$  where $c_a = \sqrt{K_a/\rho_a}$ is the speed of sound in air and $\rho_a$ is the density.   In order to quantify  the low frequency, or equivalently quasi-static,  response consider  the two  plates   subject to static pressure $p_0$ (positive or negative). The resulting   change in thickness of the air gap    is
\begin{equation}
\Delta L_a = \frac{d_a}{K_a} p_0  .
\label{DVofAir}
\end{equation}

The    flex-layer and air gap have equal compliance (or it's inverse, stiffness) if Eqs.\  (\ref{DVofFlex}) and (\ref{DVofAir}) agree, i.e.\ if 
\beq{73}
\frac{K_a} {d_a}= \frac {15}8 \frac{h^3}{b^4} \frac{E_p}{(1-\nu^2)} .
\eeq
This can be considered a constraint on the three lengths $b$, $h$ and $d_a$ for a given plate material.   We assume the plates are Aluminum ($E_p = 70$ GPa, $\nu = 0.334$), and using $K_a = 0.134$ MPa  implies the relation $b = 32.40\, (h^3 d_a)^{1/4}$. 
Assuming  plates of thickness $h=1$ mm, then for   different value of $d_a$ (0.5mm, 1 mm, 2 mm), $b$  is 2.72 cm, 3.24 cm, and 3.85 cm, respectively.

\begin{figure}[H]
    \centering
    \includegraphics[width=0.5\textwidth]{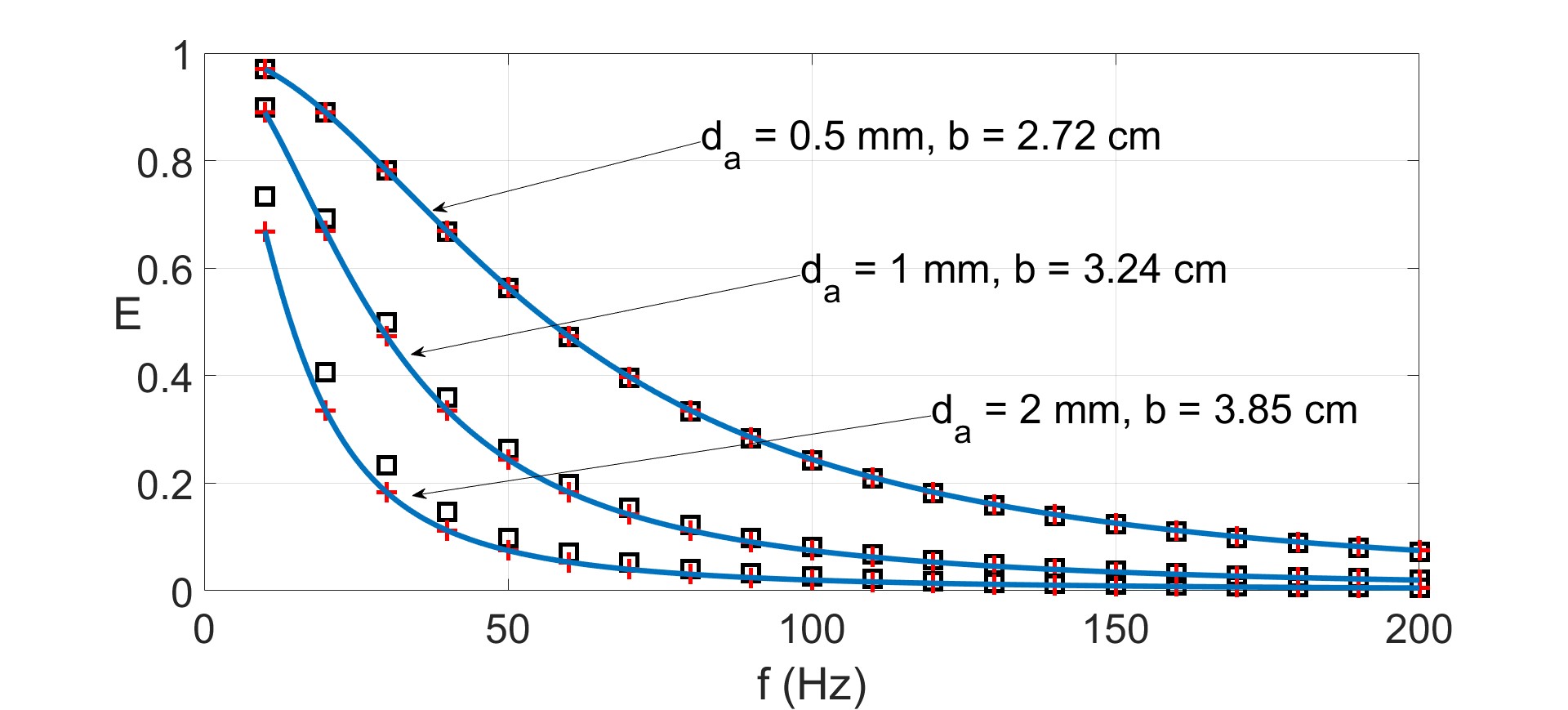}
    \caption{Transmitted energy ratio $E=|T|^2$ vs frequency for   normal incidence.   Solid lines are obtained using Eq.\ (\ref{99-}), while $\Box$ and $+$ are obtained using FEM (COMSOL) for the air gap model and flex-layer model, respectively. The plates are Aluminum  with $h=1$ mm, and the rib half-spacing $b$ is found using Eq.\ \eqref{73}.
    \rev{The non-dimensional acoustic wavenumber $kb$ where $k=\omega /c$ is less than $0.033$ for the cases shown, implying that the pressure variation along the surface is minimal for non-normal incidence.  The curves are therefore independent of the angle of incidence.}  }
    \label{fig3}
\end{figure}

The equivalence of the flex-layer and the air gap is demonstrated in Fig.\  \ref{fig3} which shows the fractional energy transmission  of both systems at low frequencies for normal incidence.  Also shown is the transmitted energy across a spring layer of stiffness $\kappa$, for which the reflection and transmission coefficients are:
\beq{99-}
R = \frac{ \frac{\ii \omega}{\kappa}}{\frac 2Z - \frac{\ii \omega}{\kappa}}, 
\qquad
T = \frac{ \frac 2Z }{\frac 2Z - \frac{\ii \omega}{\kappa}}
\eeq
where $Z = \rho c$ and $\kappa = K_a/d_a$  for the air gap.   Alternatively, $\kappa$ can be related to the flex-layer parameters via Eq.\ \eqref{73}.
This indicates that the quasi-static hypothesis and analogy with the air gap is valid at low frequencies, and that the 
reflection and trasmission depends only on the effective stiffness $\kappa$. 

We next describe an analytical model for acoustic scattering from the flex-layer  that is valid beyond the quasi-static regime. 

\section{Full dynamic model of scattering from a flex-layer}  \label{sec3}   

\begin{figure}[H]
    \centering
    \includegraphics[width=0.25\textwidth]{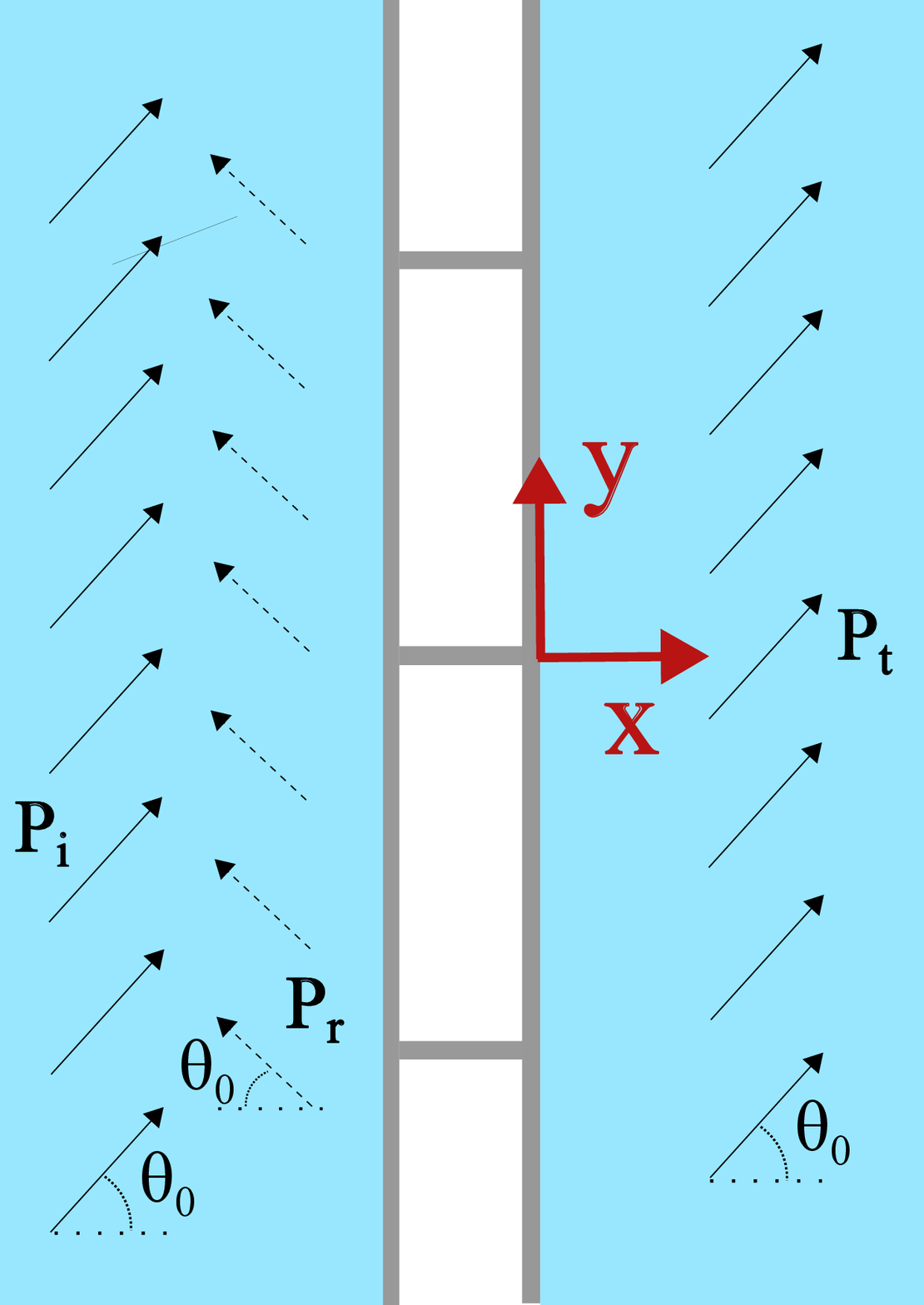}
    \caption{Incident, reflected, and transmitted waves in Water-Flex-Water model}
    \label{WFlexWR}
\end{figure}

The periodicity of the flex-layer system of Fig.\ \ref{WFlexWR}   in the $y-$direction introduces the possible generation of reflected and transmitted waves with all $y-$wavenumbers commensurate in the unit wavenumber $2\pi/d$ where $d=2b$.  We consider  oblique  plane wave incidence at angle $\theta_0$ from the normal, with $y-$wavenumber $ k\sin\theta_0$ where $k=\omega /c$.  This can then give rise to waves with wavenumbers in the $y$ and $x$ directions, respectively, 
\beq{7=17}
 k_m = k\sin\theta_0 + 2\pi \frac md, 
\ \ \
 (k_x)_m =\big( k^2 - k_m^2 \big)^{1/2}, 
\eeq
for all $m\in \mathbb{Z}$ with the square root either positive real or positive imaginary.  If the flex-layer interacts with another scatterer, e.g. another flex-layer or a material interface, the resulting multiple scattering will involve  incidence on the flex-layer by all $y-$wavenumbers.  With that purpose in mind  - see Section \ref{sec5} - we   formulate the problem in a general sense of a single flex-layer incident by $y-$wavenumber $k_n$ for some specific  $n$.

\begin{figure}[H]
    \centering
    \subfigure[\ Symmetric]{\includegraphics[width=0.2\textwidth]{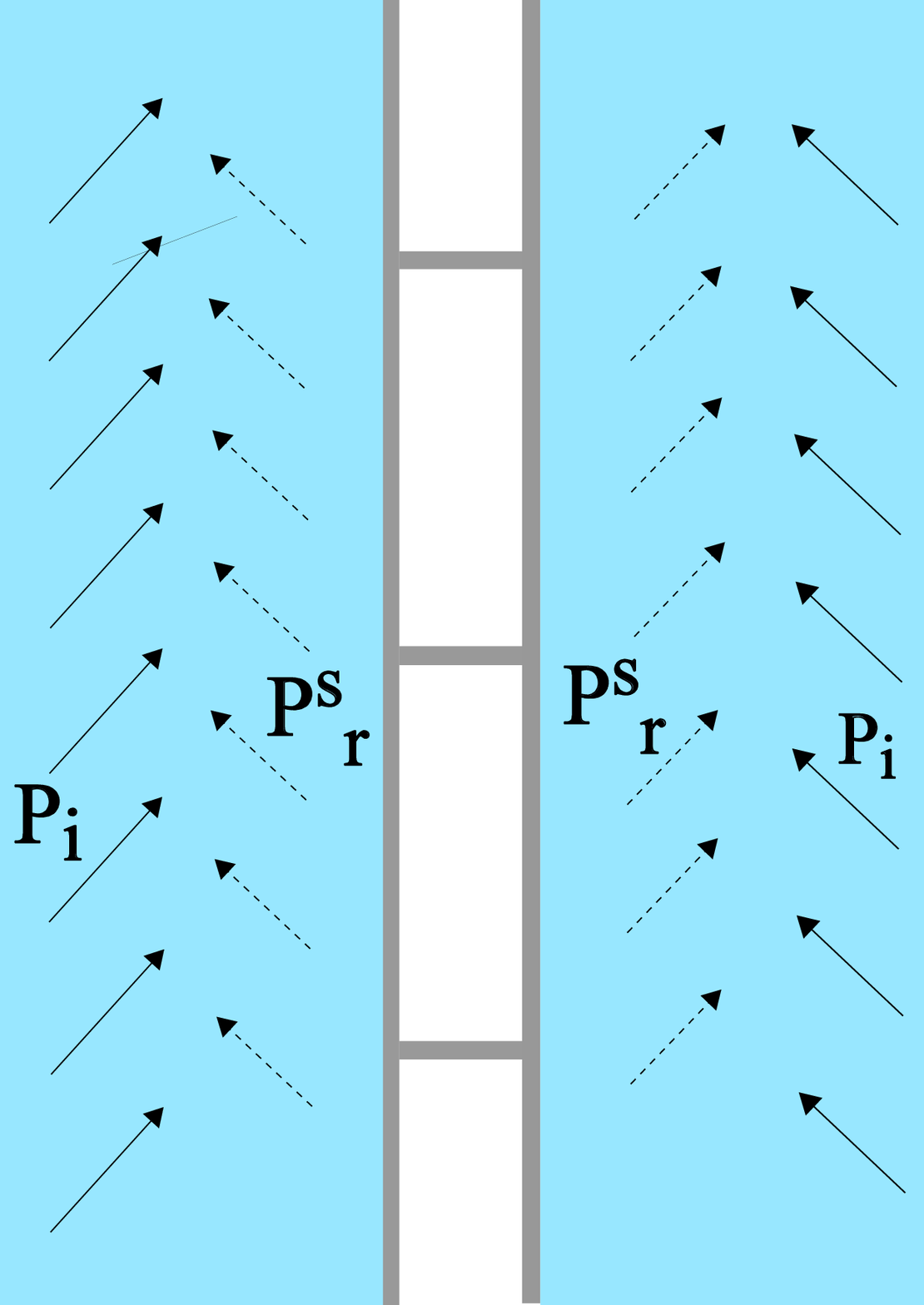}}
    \hspace{0.01\textwidth}
    \subfigure[\ Antisymmetric]{\includegraphics[width=0.2\textwidth]{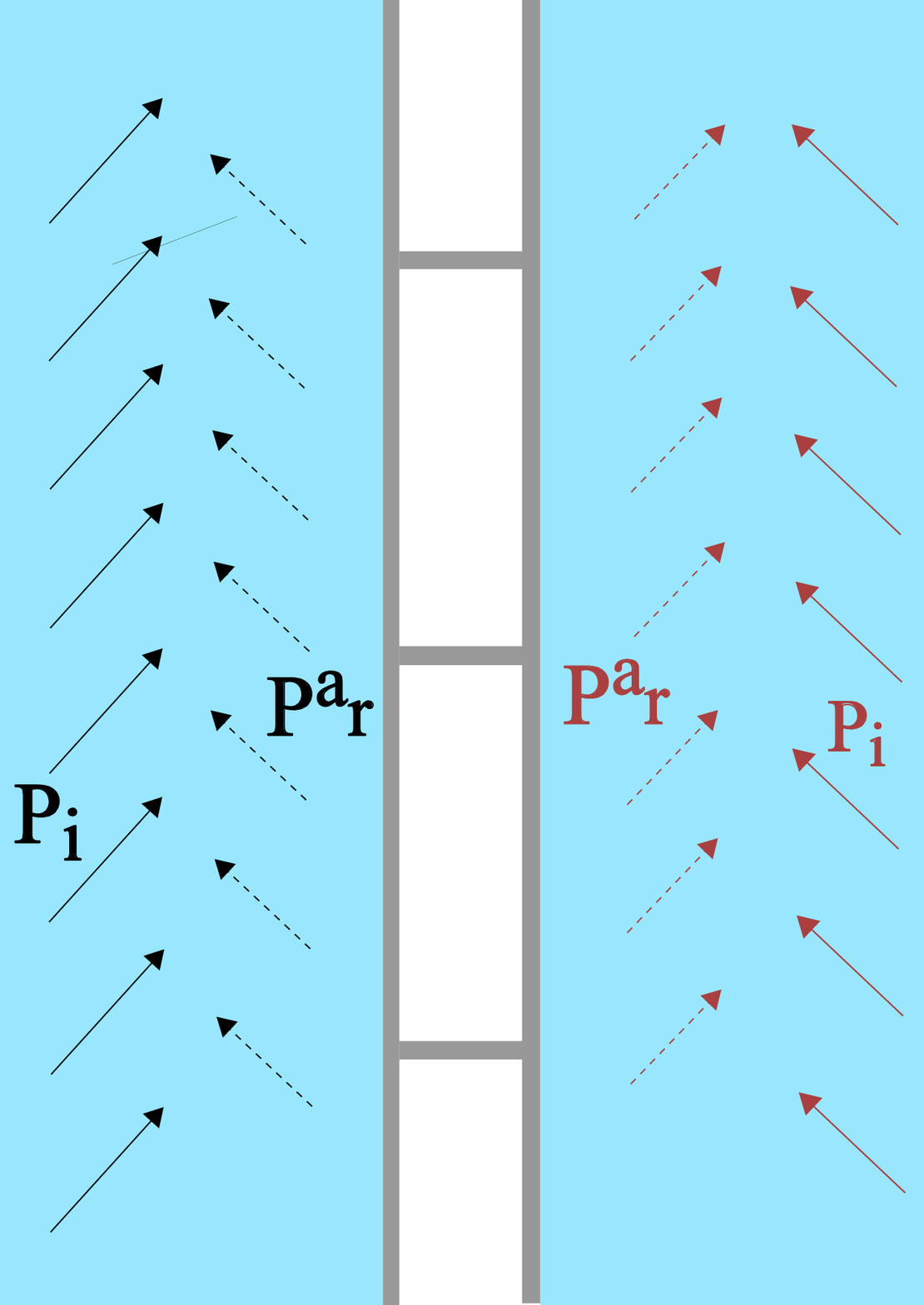}} 
    \caption{The separate symmetric and antisymmetric scattering problems.  Red indicates negative.}
    \label{WFlexWRsRas}
\end{figure}

\rev{ The problem as depicted in Fig. \ref{WFlexWR}  contains incident, reflected, and transmitted waves. The solution is obtained by splitting the scattering problem  into  distinct cases: symmetric and antisymmetric, see Fig. \ref{WFlexWRsRas}.
}

\subsection{Symmetric  scattering from  a single flex-layer} \label{SymmetricSection}

The total acoustic pressure is assumed to comprise an incident wave of $y-$wavenumber $k_n$: 
\bal{7=04}
p(x,y) =&   p_0 \, 
 \big[ e^{\ii \big( -(k_x)_n |x| + k_n y \big)   }
 \notag \\ &
+  \sum_{m=-\infty}^\infty R_{mn}^{(s)}  
              e^{\ii \big( (k_x)_m |x| + k_m y \big)   }   \big] 
\eal
where $R_{mn}^{(s)}$ are the symmetric scattering matrix elements. 
We only need to consider scattering in either $x>0$ or $x<0$ from a single periodically constrained plate.  We look at  the $x>0$ problem.  
The setup is a plate on the $y-$axis, fluid  in $x>0$, and $(2b=) \,d-$periodic  normal force constraints on the plate at $y_m= md$, $m=0, \pm 1, \pm 2, \ldots$.  The following is motivated by Stepanishen's fundamental solution for plane wave incidence on an infinite plate with periodic stiffeners \cite{Stepanishen1978}. 

The total acoustic pressure of Eq.\ \eqref{7=04} is expressed as
\beq{7=1}
p(x,y) = p_0 e^{\ii(-(k_x)_n x +k_n y)}
+        p_0 e^{\ii((k_x)_n x +k_n y)} + p_s(x,y) 
\eeq
so that the incident and rigidly reflected terms give zero normal velocity  on the plate.  The plate normal velocity, $v(y) = v_x(0,y)$ is  related to the additional pressure $p_s$  by the momentum equilibrium equation in the \rev{$x-$}direction: $\ii \omega \rho v(y) = \frac{\partial p_s}{\partial x}(0,y)$.  Introducing the $y-$ transform, 
\beq{7=2}
\begin{aligned}
\hat V(k_y) &= \int_{-\infty}^\infty v(y)  e^{-\ii k_y y} \dd y, 
\\
v(y) &= \frac 1{2\pi}  \int_{-\infty}^\infty \hat V(k_y)  e^{\ii k_y y} \dd k_y ,
\end{aligned}
\eeq
it follows that the additional scattered pressure is related to the normal velocity by 
\beq{7=3}
p_s(x,y) =  \frac 1{2\pi}  \int_{-\infty}^\infty \hat Z_f(k_y) \hat V(k_y)  
e^{\ii ( \sqrt{k^2-k_y^2}\, x + k_y y )}\, \dd k_y
\eeq
where  $\hat Z_f$ is a fluid impedance 
\beq{7=4}
\hat Z_f(k_y)= \rho \omega / \big( k^2-k_y^2\big)^{1/2} .
\eeq

The plate displacement in the $x-$direction, $w(y) = (-\ii \omega)^{-1}v(y)$, satisfies
\bal{7=5}
    D w''''(y) - \rho_s h \omega^2 w(y) & =
- 2p_0  e^{\ii k_n y } - p_s(0,y) 
\notag \\  &
-Z_0 v(y)  \sum_{m=-\infty}^\infty \delta(y-y_m) .
\eal
 This assumes  Kirchhoff plate theory; the analysis for the alternative Mindlin plate model is given in Appendix \ref{appb}.
The first two forcing terms in Eq.\ \eqref{7=5}  are from the fluid pressure and the final term in \eqref{7=5} represents the forcing from the rib constraints, with impedance $Z_0$ assumed to be the same for each rib.  Each is of length $l_\text{r}$, thickness $h_\text{r}$, 
with material properties  $E_\text{r}$ and $\rho_\text{r}$,  and \cite{rao2019vibration}
\beq{z0rs}
Z_0 = \ii \rho_\text{r}  c_\text{r} h_\text{r}  \cot \frac{\omega l_\text{r}}{2 c_\text{r} } 
\eeq
where $c_\text{r} = \big( E_\text{r} /\rho_\text{r} \big)^{1/2}$.
Taking the $k_y$ transform of \eqref{7=5} yields
\beq{7=10}
\begin{aligned}
\hat V(k_y) = - \hat Y(k_y)  \, \Big( 4\pi p_0 \, \delta(k_y - k_n) + 
 q(k_y)  
\Big)
\end{aligned}
\eeq
where $\hat Y$ is the compliance of the plate and fluid  in parallel, 
\beq{7=009}
\hat Y(k_y) = \Big( \hat Z_p (k_y) + \hat Z_f(k_y)\Big)^{-1}, 
\eeq
with plate impedance according to Kirchhoff theory (or by Eq.\ \eqref{7=91} using Mindlin plate theory) 
\beq{7=9}
\hat Z_p(k_y) = \frac {Dk_y^4 - \rho_s h\omega^2}{-\ii \omega} ,
\eeq
and 
\beq{5+3-}
q(k_y) = Z_0   \int_{-\infty}^\infty  v(y)    e^{-\ii k_y y}  \sum_{m=-\infty}^\infty    \delta(y-y_m) \dd y  .
\eeq

 \rev{ The Poisson Summation identity \cite{Zygmund,Evseev} }
\beq{7=11}
\sum_{m=-\infty}^\infty    \delta(y-y_m)
 = \frac 1d   \sum_{m=-\infty}^\infty e^{-\ii 2\pi m \frac yd}
\eeq
allows us to express Eq.\ \eqref{5+3-} as
\beq{7=13}
q(k_y) = \frac{Z_0}d  \sum_{m=-\infty}^\infty \hat V(k_y + 2\pi \frac md) .
\eeq
Noting that $q(k_y) $ is periodic with period $2\pi/d$ gives, using Eq.\ \eqref{7=10}, 
\beq{7=15}
q(k_y) = -4\pi p_0 \,  \hat Y(k_n)\, \hat Z(k_0)
 \sum_{m=-\infty}^\infty \delta(k_y- k_m) 
\eeq 
where  
\beq{7=236}
\hat Z(k_0) = \Big(  \frac d{Z_0}  + 
  \sum_{m=-\infty}^\infty \hat Y ( k_m ) \Big)^{-1} .
\eeq
Equations \eqref{7=3}, \eqref{7=10} and \eqref{7=15} yield
\bal{7=16}
    & p_s(x,y) = 2p_0  \hat Y(k_n)  \bigg\{ - \hat Z_f(k_n) 
 e^{\ii \big( (k_x)_n x +k_n y\big) } 
 \notag \\ & 
+ \hat Z (k_0)
\sum_{m=-\infty}^\infty \hat Z_f\big( k_m\big)  \hat Y\big( k_m\big)  \,
  e^{\ii \big( (k_x)_m x +k_m y\big) } 
\bigg\} .
\eal
\rev{Note that the dispersion relation for symmetric structure-borne waves \cite{Skelton2018} is $1/\hat Z(k_0) = 0$.  }

The total field \eqref{7=1} can now be written, using  \eqref{7=16}, as 
\bal{7=18}
    p(x,y) =& p_0  e^{\ii \big( -(k_x)_n x +k_n y\big) } 
+        p_0 R_p(k_n)\, e^{\ii \big( (k_x)_n x +k_n y\big) } 
\notag \\ &  
+ p_c(x,y)
\eal
where $R_p$ is the reflection coefficient for the  plate with no 
constraints,
\beq{7=19}
R_p(k_n) = \frac{\hat Z_p(k_n) -  \hat Z_f(k_n)  }{
\hat Z_p(k_n) + \hat  Z_f(k_n)  }
\eeq
and $p_c$ is  caused by the ribs,
\bal{7=20}
    p_c(x,y) = &p_0 \, \hat Z(k_0)  \hat Y(k_n)  
    \notag \\ & \times 
\sum_{m=-\infty}^\infty   \big( 1-  R_p(k_m) \big) \,  
  e^{\ii \big( (k_x)_m x +k_m y\big) }  .
\eal
In summary, referring to Eq.\ \eqref{7=04}, the  scattering matrix for symmetric incidence has elements  
\beq{066-}
{ R_{mn}^{(s)} =  R_p(k_m) \, \delta_{mn}  
+ \big( 1-  R_p(k_m) \big) \, 
\hat Z(k_0)   \hat Y(k_n)  . }
\eeq
\rev{ Acoustical reciprocity \cite{Nassar2020} is ensured by the symmetry relation 
$ R_{mn}^{(s)}  \, \hat Z_f(k_n) = R_{nm}^{(s)}  \, \hat Z_f(k_m) $.}

\subsection{Symmetric scalar problem: Propagating plane wave incidence}
In the special case that the incidence is a plane wave $(n=0)$ then the reflected pressure can be split into two parts: a reflected plane wave and a remainder comprising all modes $m\ne 0$.  The latter are all evanescent if the frequency is low enough.  Specifically, 
\bal{7=21}
    p(x,y) =& p_0 e^{\ii k(-x\cos \theta_0 + y \sin  \theta_0)} 
   \notag  \\
& +        p_0 R_{00}^{(s)}\, e^{\ii k(x\cos \theta_0 + y \sin  \theta_0)} +  p_\text{ev}^{(s)}(x,y)
\eal
where  the  reflection coefficient $R_{00}^{(s)}$ follows from \eqref{066-} as 
\beq{7=22}
R_{00}^{(s)} =
	 \frac{\hat Z_{p0} (k_0)  -  \hat Z_f(k_0)  }{\hat Z_{p0}(k_0)  + \hat  Z_f(k_0)  }
\eeq
with modified plate impedance (see \eqref{7=9}) 
\beq{7=23}
\hat Z_{p0} (k_0)=  \hat Z_p(k_0) + 
\Big(  \hat Z^{-1}(k_0) -\hat Y (k_0)     \Big)^{-1}, 
\eeq
and evanescent field
\bal{7=201}
    p_\text{ev}^{(s)}(x,y)= &p_0 \, \hat Z(k_0)  \hat Y(k_0)  
\notag \\  & \times 
\sum_{m \ne 0}  \big( 1-  R_p(k_m) \big) \,   e^{\ii \big( (k_x)_m x +k_m y\big) }  .
\eal

If the frequency is low enough that all of the  $ (k_x)_m$ are imaginary except for $m=0$ $\big( k< \frac {2\pi}d (1+\sin|\theta_0|)^{-1} \big)$ then $ p_\text{ev}(x,y)$ is evanescent in the $x-$direction.  The only acoustic energy that radiates to infinity comes from the plane wave. Total energy is conserved if $|R_{00}|=1$, which  is the case only if $\text{Re}\, Z_0 =0$. 
In the limit of closely spaced ribs $(d\to 0)$  to leading order   the sum in \eqref{7=23} is ignorable,  $\hat Z_{p0}  \approx\hat Z_{p}(k_0) +   Z_0/d$, and   \eqref{7=22} agrees with \cite[Eq.\ (32)]{Stepanishen1978}.

\subsection{Antisymmetric problem } \label{AntrisymmetricSection}

The total acoustic pressure is assumed to be 
\bal{7=24}
p(x,y) =&   -\text{sgn}(x)\, p_0 
 \big[ e^{-\ii \big( (k_x)_n |x| + k_n y \big)   } 
 \notag \\  & \qquad 
+   \sum_{m=-\infty}^\infty R_{mn}^{(a)} 
              e^{\ii \big( (k_x)_m |x| + k_m y \big)   } \big] .
\eal

\rev{\subsubsection{An approximate mass dominated solution} }

The flex-layer moves in the $x-$direction as one with velocity 
$v_x(0,y) = (\ii \omega \rho)^{-1} p_{,x}(0,y)$. 
The force-acceleration relation can be approximated as   
\beq{7=25}
-\ii \omega m_t v_x(0,y) \approx p(-0,y) - p(+0,y),
\eeq
where \rev{the total mass density $m_t$ incorporates the plate and rib masses in a cell-averaged sense, $m_t = 2\rho_s h +   \rho_\text{r}  h_\text{r}   l_\text{r} /d $
and $\rho_\text{r}$,  $h_\text{r} $ and $ l_\text{r}$ are density, thickness and length of the connecting  ribs.} 
The antisymmetric scattering matrix is therefore diagonal with elements  
\beq{7=256}
 { R_{mn}^{(a)}  \approx  
 \frac{\hat Z_\text{mass}  - \hat Z_f(k_m) }
     {\hat Z_\text{mass}  + \hat Z_f(k_m) }
 \,		\delta_{mn} }
\eeq
with  mass-like impedance $ \hat Z_\text{mass} = -\frac{ \ii }2 \omega {m_t}. $

\rev{\subsubsection{Exact solution}
As in the symmetric problem we only consider $x>0$, and express  
the total acoustic pressure in the form (see \eqref{7=1})
\beq{7=-1}
p(x,y) = -p_0 e^{\ii(-(k_x)_n x +k_n y)}
         -p_0 e^{\ii((k_x)_n x +k_n y)} + p_s(x,y) 
\eeq
with additional scattered pressure \eqref{7=3}.  
The plate displacement   satisfies (see \eqref{7=5})
\bal{7=-5}
  & D w''''(y) - \rho_s h \omega^2 w(y) =
 2p_0  e^{\ii k_n y } - p_s(0,y) 
 \notag \\ & \qquad \qquad \qquad 
-Z_\text{0mass}\, v(y) \, \sum_{m=-\infty}^\infty \delta(y-y_m) .
\eal
where $Z_\text{0mass}$ is the rib mass impedance  
\beq{3+6}
Z_\text{0mass} = - \ii \rho_\text{r}  c_\text{r} h_\text{r}  \tan \frac{\omega l_\text{r}}{2 c_\text{r} } .
\eeq
The solution follows in the same way as for the symmetric problem.  Based on eqs.\ \eqref{7=5}, \eqref{066-} and \eqref{7=24} we  find that the exact antisymmetric scattering matrix has elements  
\beq{7=+6}
{ R_{mn}^{(a)} =  R_p(k_m) \, \delta_{mn}  
+ \big( 1-  R_p(k_m) \big) \, 
\hat Z_\text{mass}(k_0)   \hat Y(k_n)   }
\eeq
where, see \eqref{7=236}, 
\beq{7=29}
\hat Z_\text{mass} (k_0) = \Big(  \frac d{Z_\text{0mass}}  + 
  \sum_{m=-\infty}^\infty \hat Y\big( k_m\big) \Big)^{-1} .
\eeq
It may be shown that \eqref{7=+6} reduces to the mass dominated approximation \eqref{7=256} as $\omega \to 0$. 
Also,  the dispersion relation for anti-symmetric structure-borne waves  is $1/\hat Z_\text{mass} (k_0) = 0$.  }

\subsection{Scattering of an incident wave from a flex-layer}
A  wave is incident from $x<0$ only, with total acoustic pressure  
\beq{7=207}
p(x,y) =  p_0\times \begin{cases}
 \big[ e^{\ii \big( (k_x)_n x + k_n y \big)   }   &
 \\ 
 +  \sum_{m=-\infty}^\infty R_{mn}
              e^{\ii \big(- (k_x)_m x + k_m y \big)   }  \big], & x<0,
\\
    \sum_{m=-\infty}^\infty T_{mn} 
              e^{\ii \big( (k_x)_m x + k_m y \big)   }  , & x>0 .
\end{cases}
\eeq
Combining the two separate solutions in Eq.\ \eqref{7=04} and  \eqref{7=24}
the reflection and transmission matrices follow as  
\beq{7=28}
R_{mn} = \frac 12 \big(R_{mn}^{(s)} +R_{mn}^{(a)}\big) . \quad   
T_{mn} = \frac 12 \big(R_{mn}^{(s)} -R_{mn}^{(a)}\big). 
\eeq
\rev{Hence, 
\beq{7-0}
\begin{aligned} 
R_{mn} =&   \frac 12 \big(
\hat Z(k_0) +\hat Z_\text{mass}(k_0) \big) \, \big( 1-  R_p(k_m) \big) \, \hat Y(k_n)  
\\  & 
+ R_p(k_m) \, \delta_{mn} ,
\\
T_{mn} =& \frac 12 \big(
\hat Z(k_0) -\hat Z_\text{mass}(k_0) \big) \, \big( 1-  R_p(k_m) \big) \, \hat Y(k_n) .
\end{aligned} 
\eeq 
}

The above general solution is valid for propagating and evanescent incident waves. In the special case of a single propagating incident plane wave at low frequency, we have 
\beq{7=27}
p(x,y) =    \begin{cases}
 p_0 e^{\ii k(x\cos \theta_0 +y \sin  \theta_0)} & 
 \\ 
 +  R_{00} p_0 e^{\ii k(-x\cos \theta_0 + y \sin  \theta_0)} + p_\text{ev}(x,y), & x<0,
\\
T_{00}        p_0 e^{\ii k(x\cos \theta_0 + y \sin  \theta_0)}  + p_\text{ev}(x,y) , & x>0
\end{cases}
\eeq
where 
\beq{7=282}
p_\text{ev}(x,y) 
=   p_0\,  \sum_{m \ne 0}  e^{\ii k_m y}\, \times
\begin{cases}
  R_{mn}               e^{- \ii  (k_x)_m x   }  , & x<0,
\\
    T_{mn}               e^{ \ii  (k_x)_m x   } , & x>0 .
\end{cases}
\eeq
Energy conservation requires that $|R_{00}|^2 + |T_{00}|^2 =1$. 

\section{Verification and Analysis of the Flex-layer Model}    \label{sec4}   
\begin{figure}[H] 
    \centering
    \includegraphics[width=0.5\textwidth]{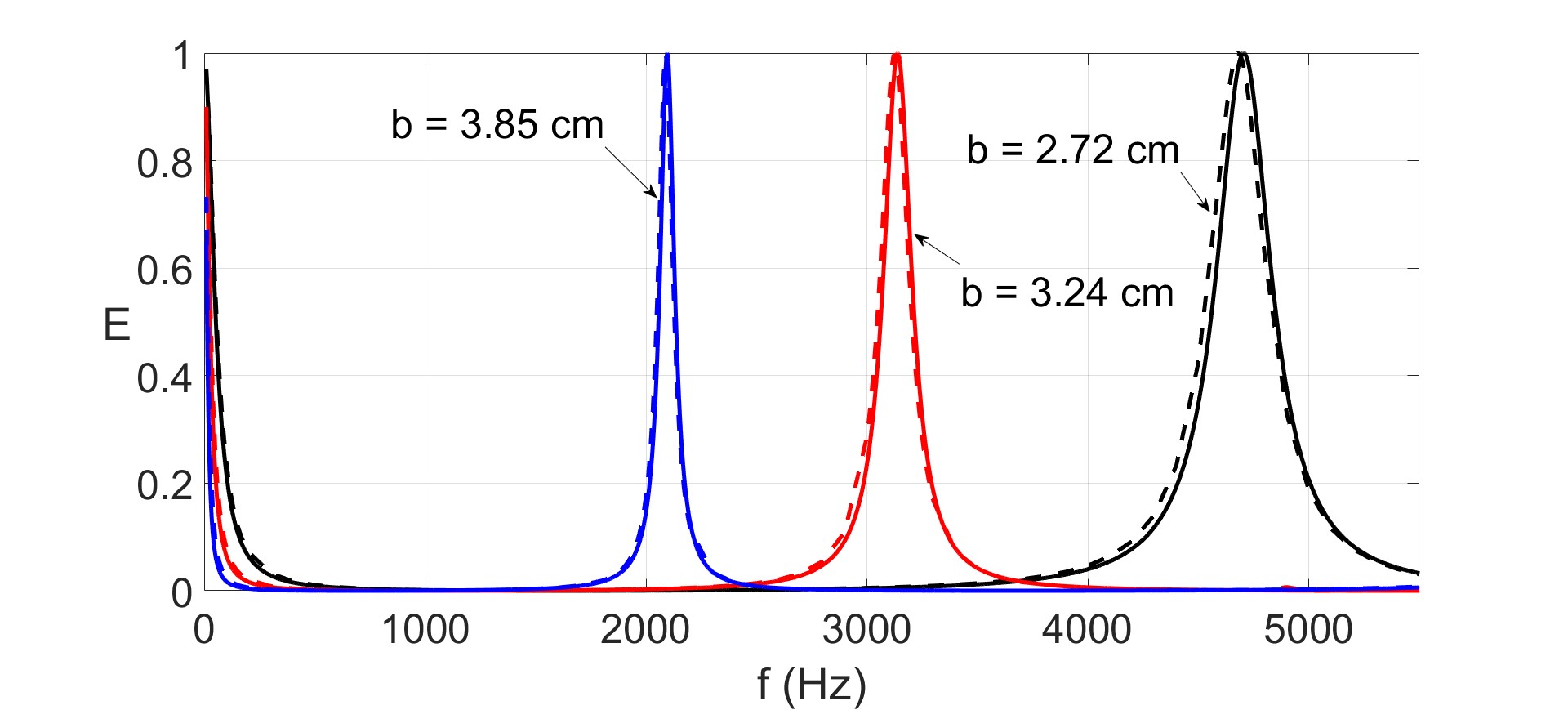}
    \caption{The transmitted acoustic energy $E$ vs frequency for three different  flex-layers each with $1$ mm Aluminum plates.  Solid lines and dashed lines are obtained using Eq.\ \eqref{7=28} and COMSOL, respectively.}
    \label{n1E}
\end{figure} 

The transmitted acoustic energy $E =  |T_{00}|^2$ plotted in Fig.\ \ref{n1E} shows that the theoretical result agrees with full FEM simulation.  Three different  flex-layers are considered with  rib spacing parameter $b$ corresponding to air gaps of width $d_a=0.5$, $1$, and $2$ mm, see Fig.\ \ref{fig3}.  The plots in Fig.\ \ref{n1E} again verify the equivalence of the flex-layer to the air gaps in the low frequency range. 
\begin{figure}[H]
    \centering
    \subfigure[Real part]{\includegraphics[width=0.5\textwidth]{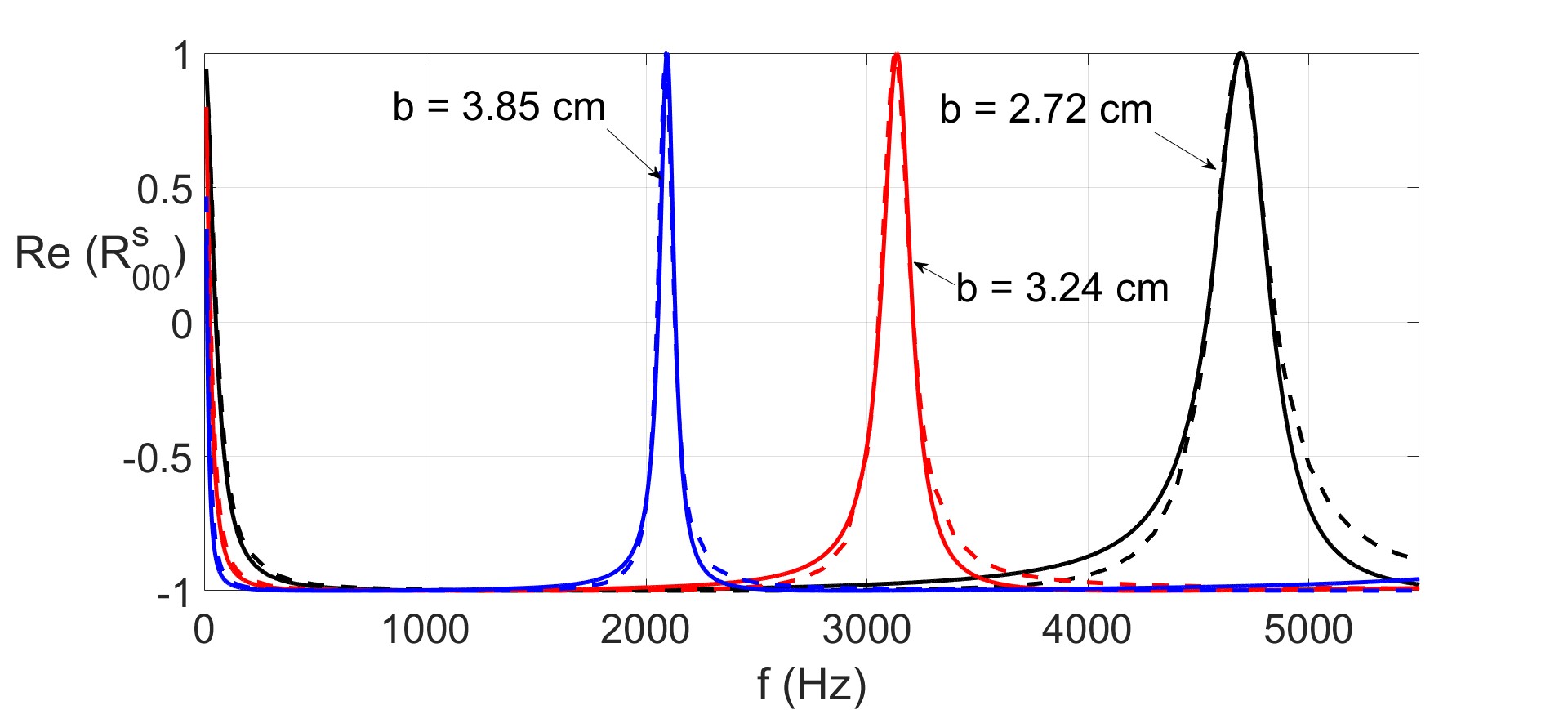}} 
    \hspace{-.15in}
    \subfigure[ Imaginary part]{\includegraphics[width=0.5\textwidth]{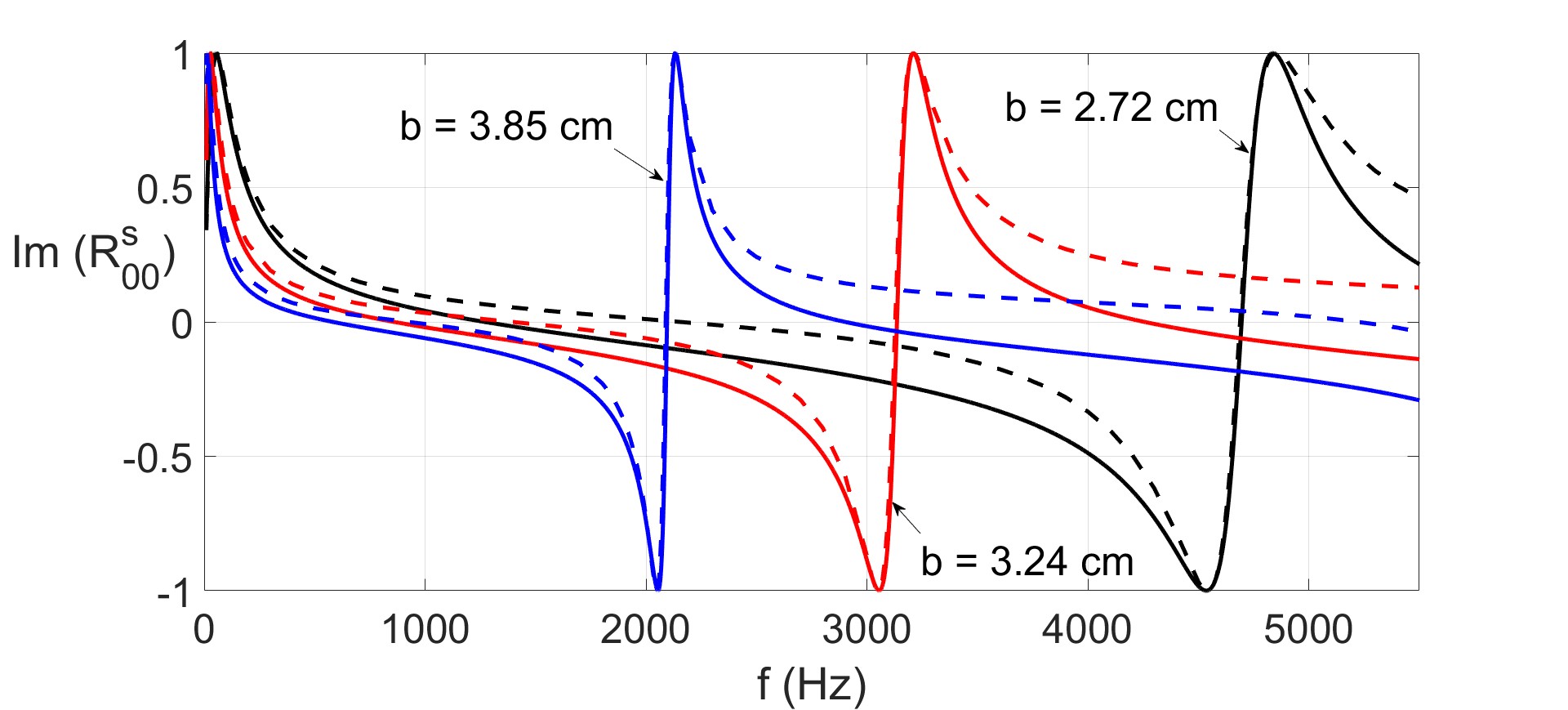}} 
    \caption{Comparison of the symmetric reflection coefficient $R_{00}^{(s)}$ vs frequency for the flex-layer with $1$ mm plates. Solid lines and dashed lines are obtained using Eq.\ \eqref{066-} and COMSOL, respectively. }
    \label{ReImRs}
\end{figure}
 The equivalence is true up to at least $1$ kHz.   However,  Fig.\ \ref{n1E} illustrates a new phenomenon of total transmission at frequencies in the range $2$ to $5$ kHz.  These values are much less than the lowest frequencies expected for total transmission through the air-gaps, i.e.\ at $f = \frac {c_a}{2d_a}$ which are approximately  twenty times as large as those of Fig.\ \ref{n1E}.  

The transmission frequencies of  Fig.\ \ref{n1E} can be understood by noting that the antisymmetric reflection coefficient $R^{(a)}_{00}$ of Eq.\ \eqref{7=256} is approximately $-1$, indicating that $E$ of  Eq.\ \eqref{7=28} equals $1$ when $R^{(s)}_{00} =1$, or equivalently  $R_{00} =0$.   This is borne out by comparing the results of  Fig.\ \ref{ReImRs} with those of  Fig.\ \ref{n1E}. 

Referring to Eq.\ \eqref{7=22}, the coefficient $R^{(s)}_{00} =1$
when $\hat Z_{p0} \to \infty$.  Equation \eqref{7=23} implies this occurs when $ \sum_{m\ne 0} \hat Y\big( k_m\big) =0$.  Assuming normal incidence $(\theta_0=0)$ this can be expressed  as
\beq{3+5}
\sum_{m=1}^\infty
\bigg( m^4 \xi^4 - \Omega^2 - 
\frac{\epsilon \Omega^2}{\sqrt{m^2 \xi^2 -\Omega^2} }
\bigg)^{-1}= 0
\eeq
with  non-dimensional frequency $\Omega = \omega /\omega_c$ and 
\beq{3=55}
\omega_c = c^2 \sqrt{\frac{\rho_s h}D},
\qquad
\epsilon = \frac{\rho c}{\rho_s h \omega_c}, 
\qquad
\xi = \frac{2\pi c}{d \,\omega_c}.
\eeq
Here $\omega_c$ is the coincidence frequency  at which
the phase velocity of the bare plate flexural wave coincides with $c$, and  $\epsilon$ is a common  non-dimensional  measure of  fluid loading \cite{Junger86}.   
\rev{ Note that  the related sum $\sum_{m=-\infty}^{\infty}  \hat Y\big( k_m\big) $ can be asymptotically approximated  
under the  heavy fluid loading limit \cite{Skelton2018} for which $\big( \rho \omega^2/D\big)^ {\frac 15} \gg k_0$ and   
$\big( \rho \omega^2/D\big)^ {\frac 15} \gg   \big( \rho_s h \omega^2/D\big)^ {\frac 14} $.   This limit is, however, not relevant to the parameters considered here. }

\begin{figure}[H] 
    \centering
    \includegraphics[width=0.5\textwidth]{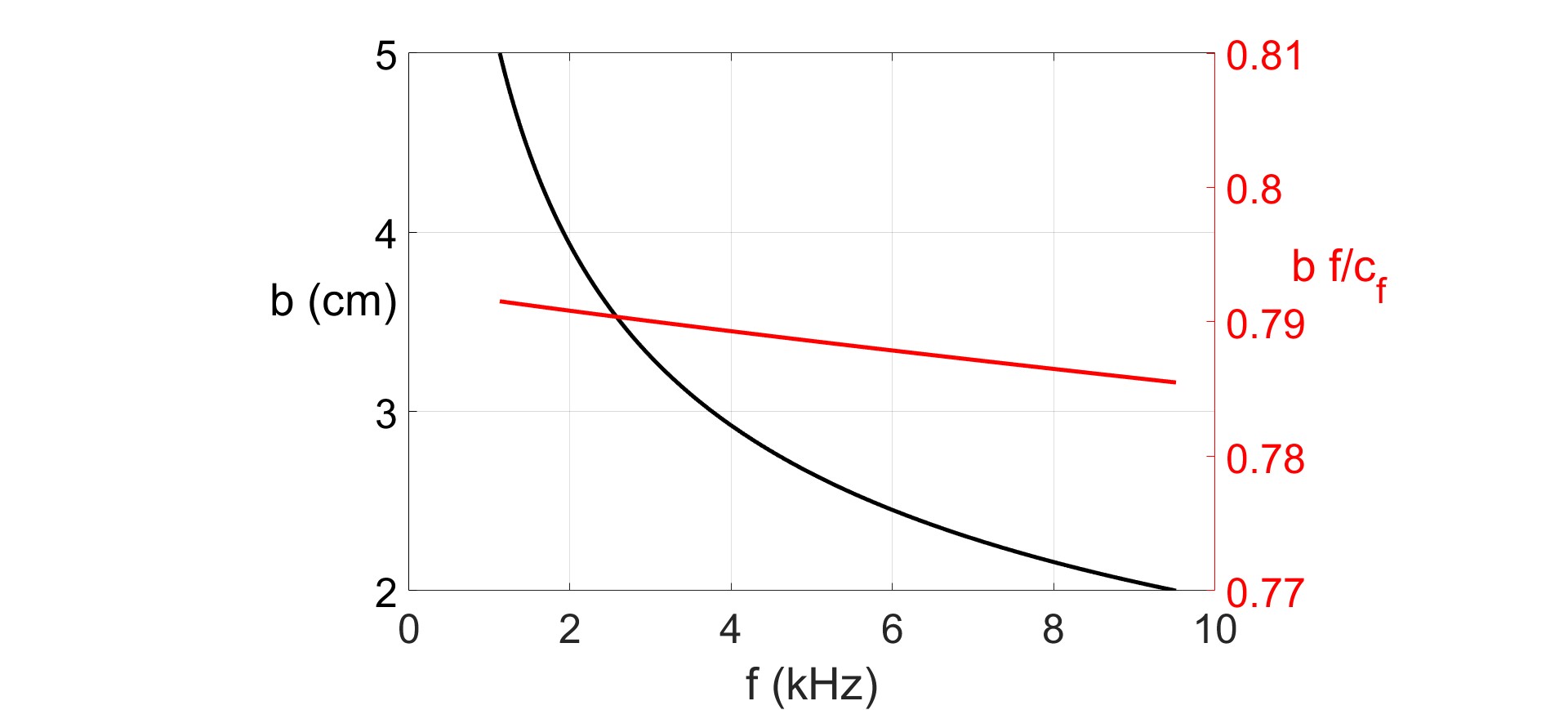}
    \caption{Two views of the full transmission frequency of the flex-layer, noted in Fig.\ \ref{n1E} for three values of the rib separation parameter $b$.  The black curve (left) shows  $b$ as a function of the full transmission frequency using Eq.\ (\ref{3+5}), \rev{i.e.\  $ \sum_{m\ne 0} \hat Y\big( k_m\big) =0$}; the red curve (right) is the non-dimensional measure $b f/c_f$ where $c_f$ is the fluid-loaded plate phase velocity obtained from Eq.\ (\ref{--1}), 
    \rev{i.e.\ $1/ \hat Y(k_y) = 0$.}  }
    \label{bvsf}
\end{figure}
\begin{figure}[H] 
    \centering
    \includegraphics[width=0.5\textwidth]{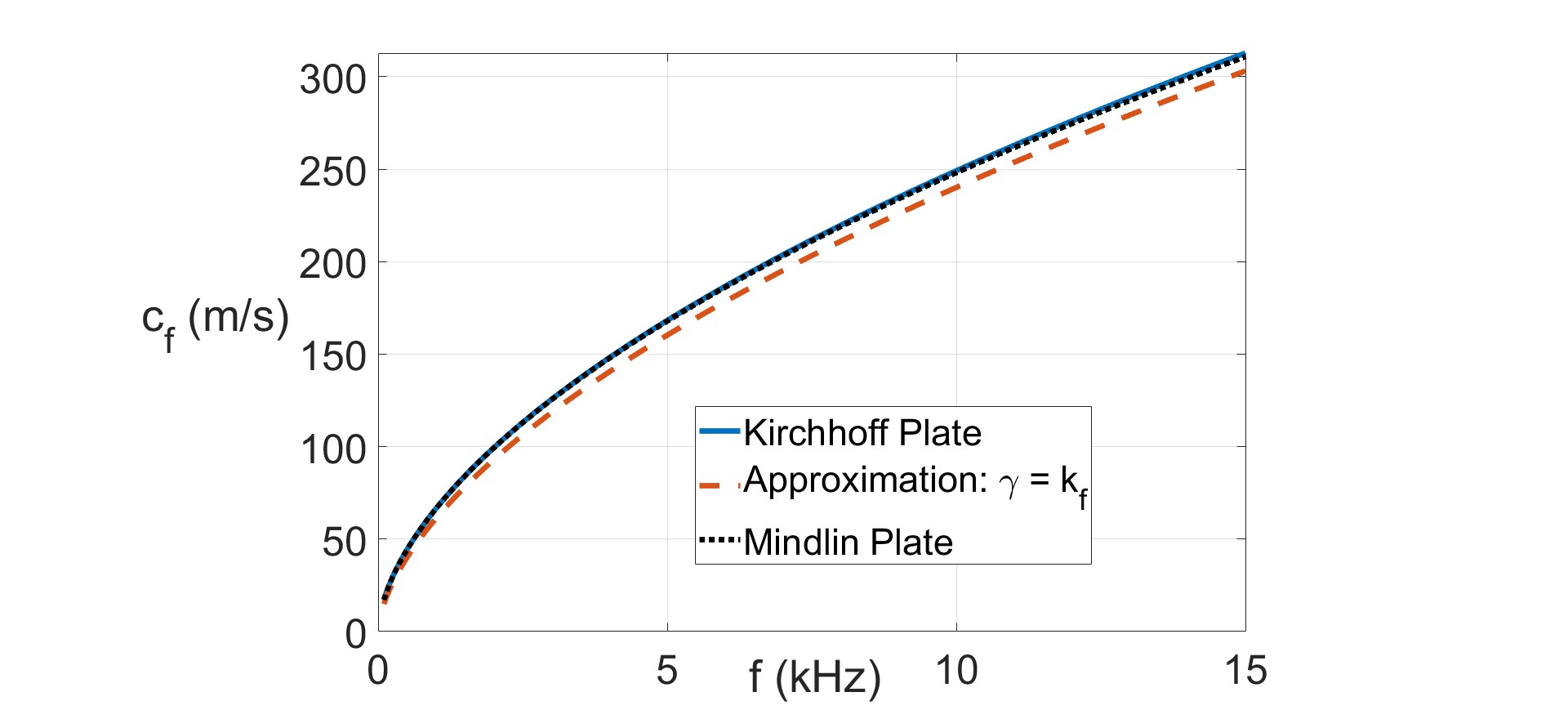}
    \caption{Flexural wave phase velocity $c_f$ vs frequency}
    \label{cfvsf}
\end{figure}

 
The physical origin of the transmission frequencies of  Fig.\ \ref{n1E} 
is actually the reverberation of fluid-loaded flexural waves, see Fig.\ref{bvsf}. 
The fluid-loaded flexural wave-number $k_y$ is the solution of 
$\hat Z_p (k_y) + \hat Z_f(k_y) = 0$, 
which  can be expressed in  terms of the flexural wave phase velocity $c_f = \omega /k_y$ as  $\tau  = \sqrt{c^2/c^2_f -1}$, the positive root of 
\cite[Eq.\ (8.8)]{Junger86}
\beq{--1}
\tau^5 + 2 \tau^3 +(1- \Omega^{-2}) \tau - \epsilon \,\Omega^{-3}=0 .
\eeq
Figure \ref{bvsf} shows that the non-dimensional rib parameter  $b f/c_f$ is approximately constant over a ten-fold range of frequency, indicating that the transmission peak is associated with flexural waves bouncing back and forth between ribs.  The small variation of  $b f/c_f$ between $1$ and $10$ kHz can be attributed to frequency dependent  interaction of the plate wave with the ribs. Numerical solution of $c_f$ in Eq.\ (\ref{--1}) versus frequency is shown in Fig.\ (\ref{cfvsf}) along with an explicit  approximate solution  ``$\gamma = k_f$" \cite[Eq.\ (8.10)]{Junger86}. 

Finally, the scattered velocity in the vicinity of the flex-layer is shown in Fig.\ \ref{fig10}
where 
\beq{0909} 
V^{sc}_x(x,y)  = \frac 1{\ii \omega \rho} \frac{\partial p_s}{\partial x}(x,y), 
 \quad
V^{sc}_y(x,y) = \frac 1{\ii \omega \rho} \frac{\partial p_s}{\partial y}(x,y).
  \eeq
Note the similarity of the velocity profiles even as the rib spacing $b$ and frequency $f$ change. 

  \begin{widetext}      \begin{minipage}{\linewidth}    
\begin{figure}[H]
    \centering
    \subfigure[]{\includegraphics[width=0.49\textwidth]{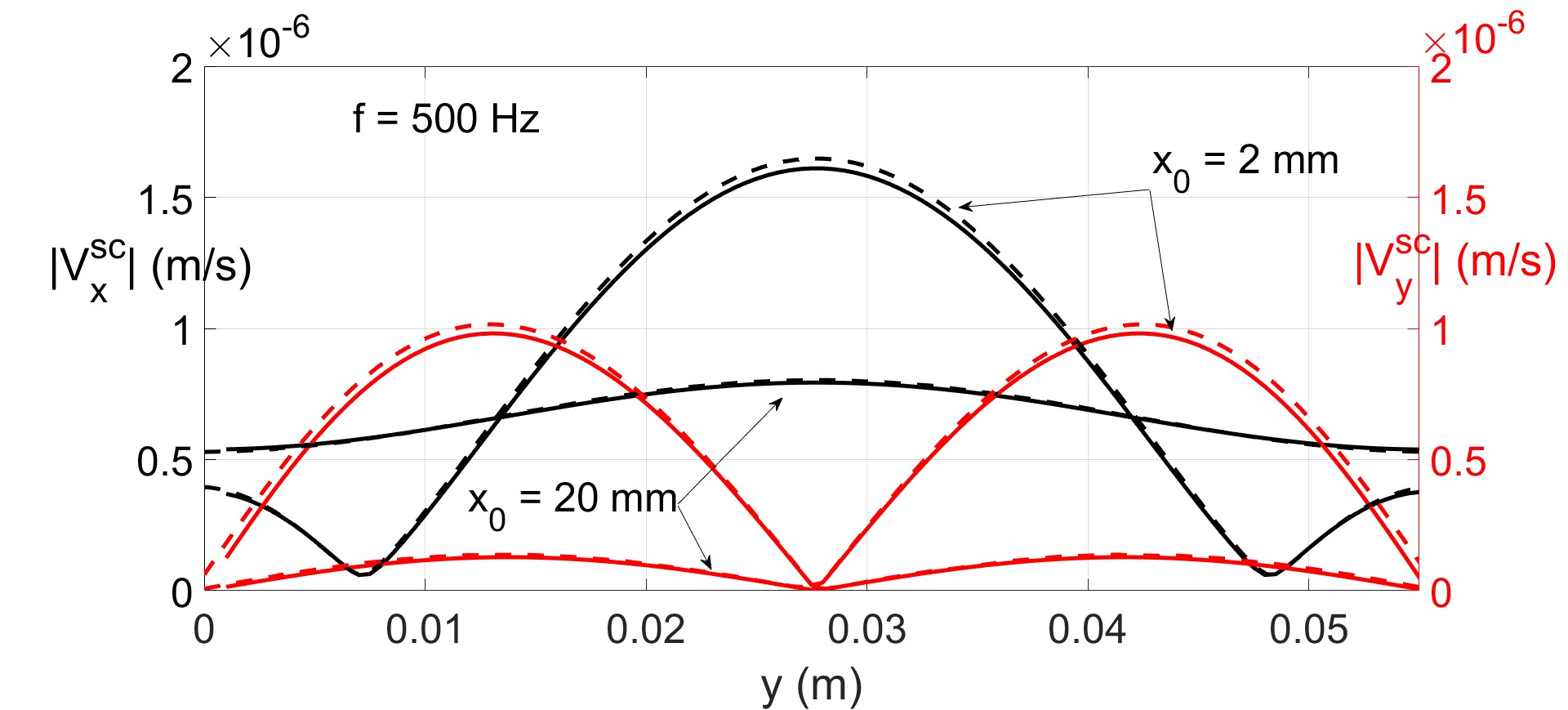}} 
    \subfigure[]{\includegraphics[width=0.49\textwidth]{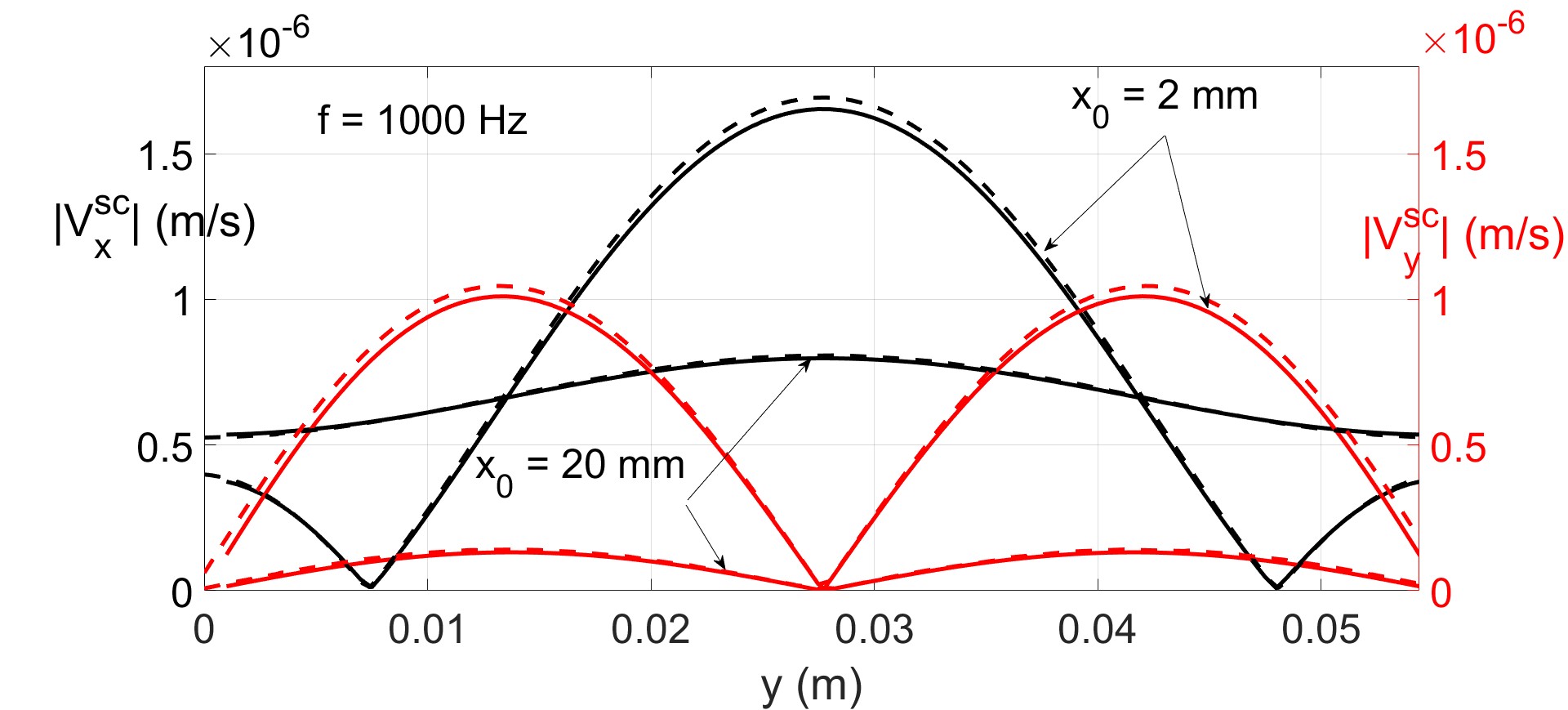}}
    \subfigure[]{\includegraphics[width=0.49\textwidth]{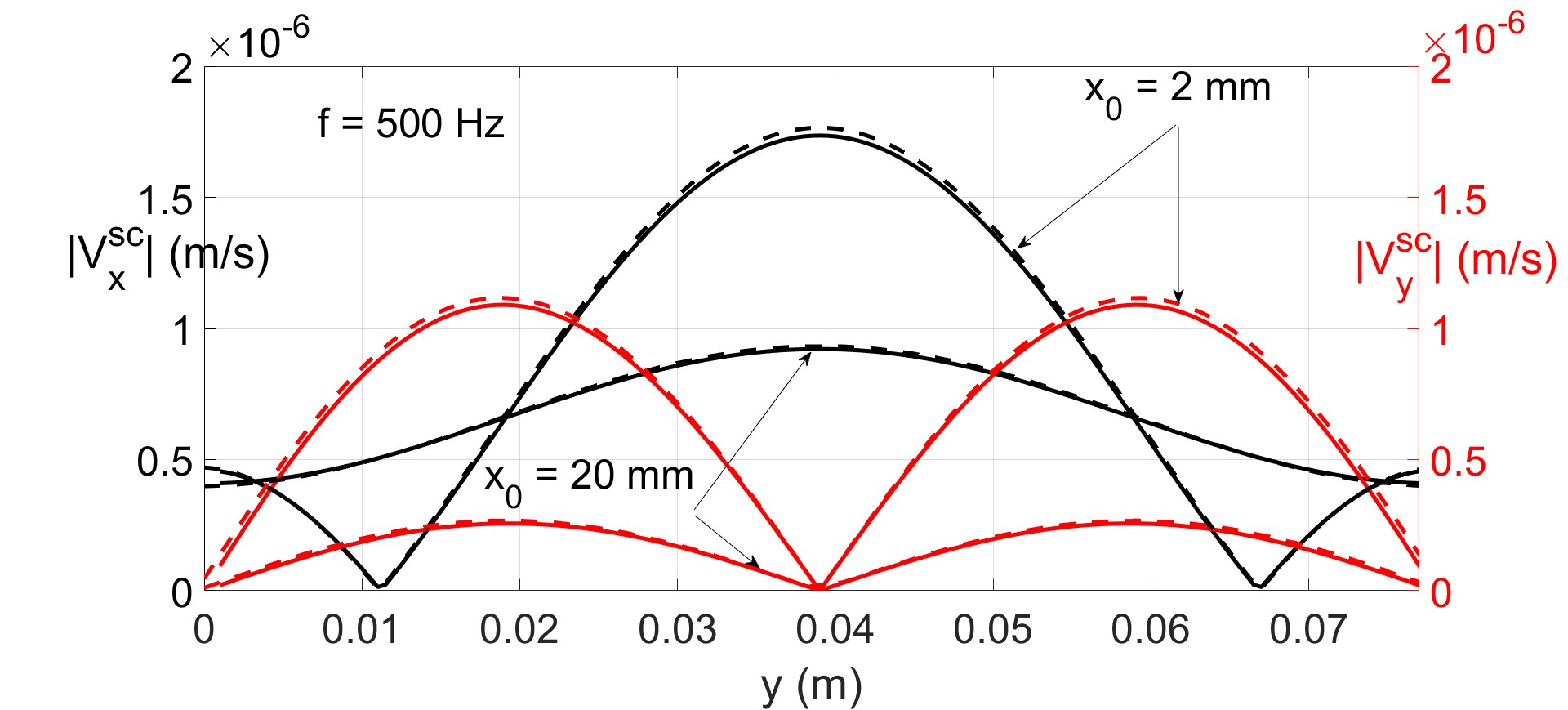}} 
    \subfigure[]{\includegraphics[width=0.49\textwidth]{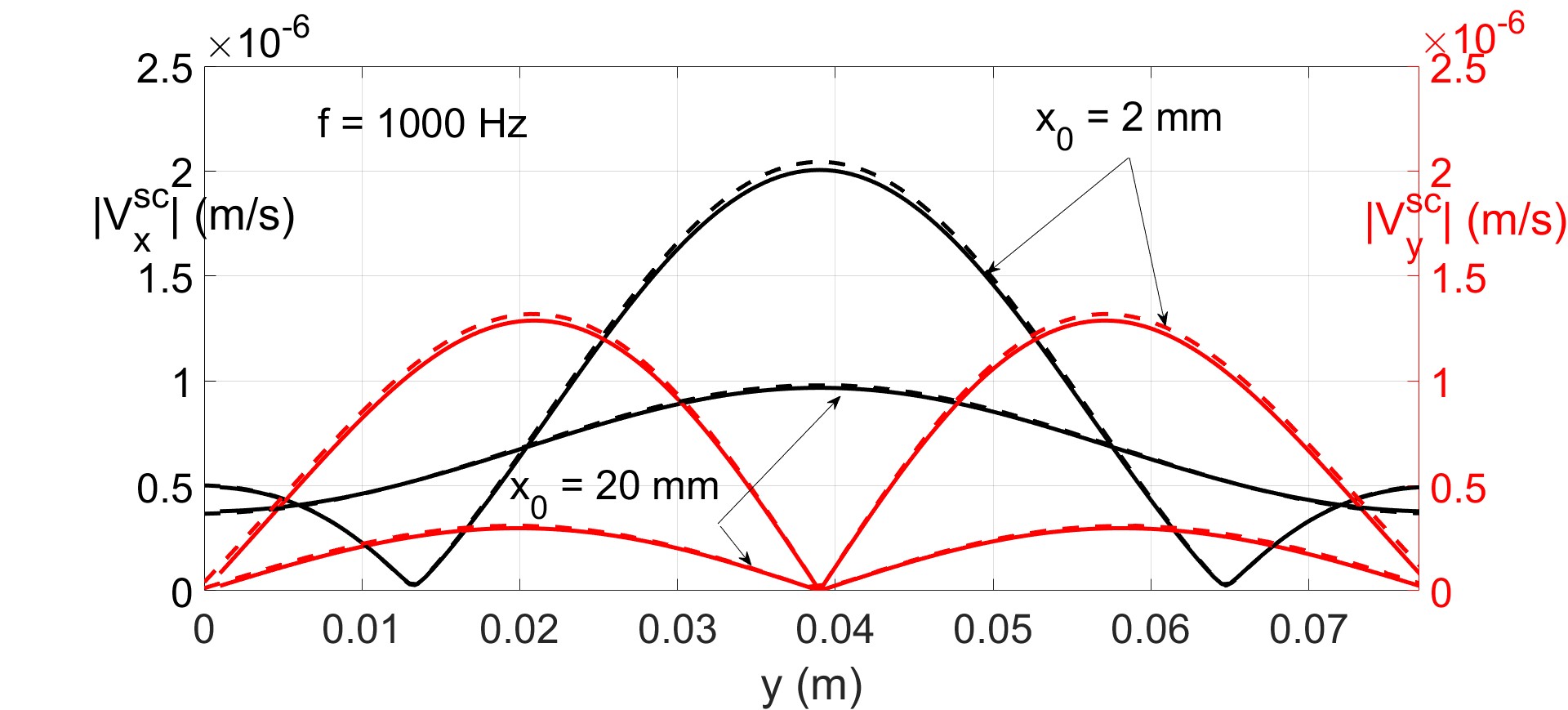}}
    \caption{Magnitude of scattered velocities in x and y directions measured at a plane  a distance of $x_0$ from the flex-layer. Solid lines and dashed lines are obtained using Eq.\ \eqref{0909} and COMSOL, respectively. $p_0 = 1$ Pa, (a,b) $b$ = 2.72 cm, (c,d) $b$ = 3.85 cm }
    \label{fig10}
\end{figure}
  \end{minipage} \end{widetext}                     

\section{Multiple flex-layers in series: $n=2$} \label{sec5}   
\begin{figure}[H]
    \centering
    \includegraphics[width=0.2\textwidth]{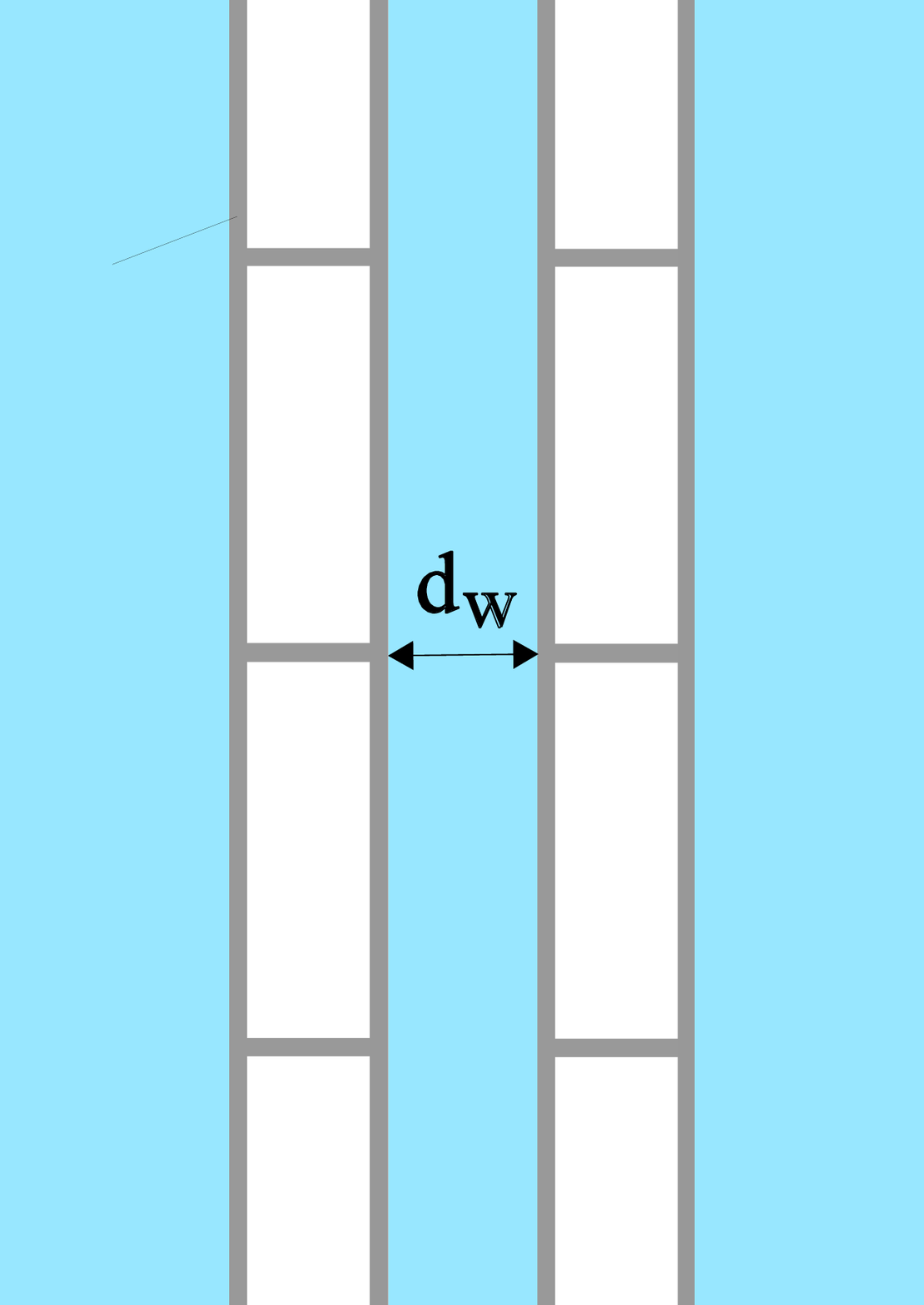}
    \caption{Two flex-layers in series separated by a water gap of width $d_w$.}
    \label{W2FlexW}
\end{figure}

Consider two flex-layers separated by a gap as shown in Fig.\ \ref{W2FlexW}.  A plane wave incident from  the left or the right will generate an infinite set of evanescent waves in the gap, which will in turn reverberate between the flex-layers.  However, the full transmission and reflection solution can be found in a semi-analytic form using the fundamental solution for the single flex-layer of Eqs.\   \eqref{7=207} and \eqref{7=28}.  The multiple scattering solution can be found by analogy with the scalar, or single wave, multiple scattering problem.  

Since the flex-layer converts a plane wave into the infinite reflected and transmitted constituents 
 we work with infinite vectors ${\bf u}^+$ and ${\bf u}_-$ corresponding to propagation in the positive and negative $x-$directions.  We consider a propagating plane  wave incident  from the left,  ${\bf u}^+_0$,  where $\{u^+_0\}_m = \delta_{m0}$.  
The wave reflected from a single flex-layer is
${\bf R} {\bf u}^+_0$, and the transmitted wave is ${\bf T} {\bf u}^+_0$ where ${\bf R}$ and  ${\bf T}$
are the infinite matrices defined by Eqs\   \eqref{7=28}. 

The  full transmission and reflection from the two flex-layers in series   follows from a ray summation approach.   The wave transmitted to the right, and the wave reflected to the left are, respectively, 
\beq{0430}\begin{aligned}
{\bf u}^+ &= {\bf T} \big( {\bf 1} + {\bf A} +  {\bf A}^2 +\ldots \big) {\bf P} {\bf T}  \, {\bf u}^+_0,
\\
{\bf u}^- &=  {\bf R} {\bf u}^+_0  + {\bf T} {\bf P}  {\bf R} 
\big( {\bf 1} + {\bf A} +  {\bf A}^2 +\ldots \big) {\bf P} {\bf T}  \, {\bf u}^+_0,
\end{aligned}
\eeq
where   ${\bf P}$ is the propagator matrix through  the water layer of thickness $d_w$, with 
$\{P\}_{mn} = e^{\ii d_w (k_x)_m} \delta_{mn}$, and 
$
{\bf A}= ( {\bf P} {\bf R} )^2$.
Hence,
\beq{43-}
\begin{aligned}
{\bf u}^+ &= {\bf T} \big( {\bf 1}- {\bf A} \big)^{-1} {\bf P} {\bf T}  \, {\bf u}^+_0,
\\
{\bf u}^- &=  {\bf R} {\bf u}^+_0  + {\bf T} {\bf P}  {\bf R} 
\big( {\bf 1}- {\bf A} \big)^{-1} {\bf P} {\bf T}  \, {\bf u}^+_0.
\end{aligned}
\eeq

\begin{widetext}      \begin{minipage}{\linewidth}    
\begin{figure}[H]
    \centering
    \subfigure[]{\includegraphics[width=0.49\textwidth]{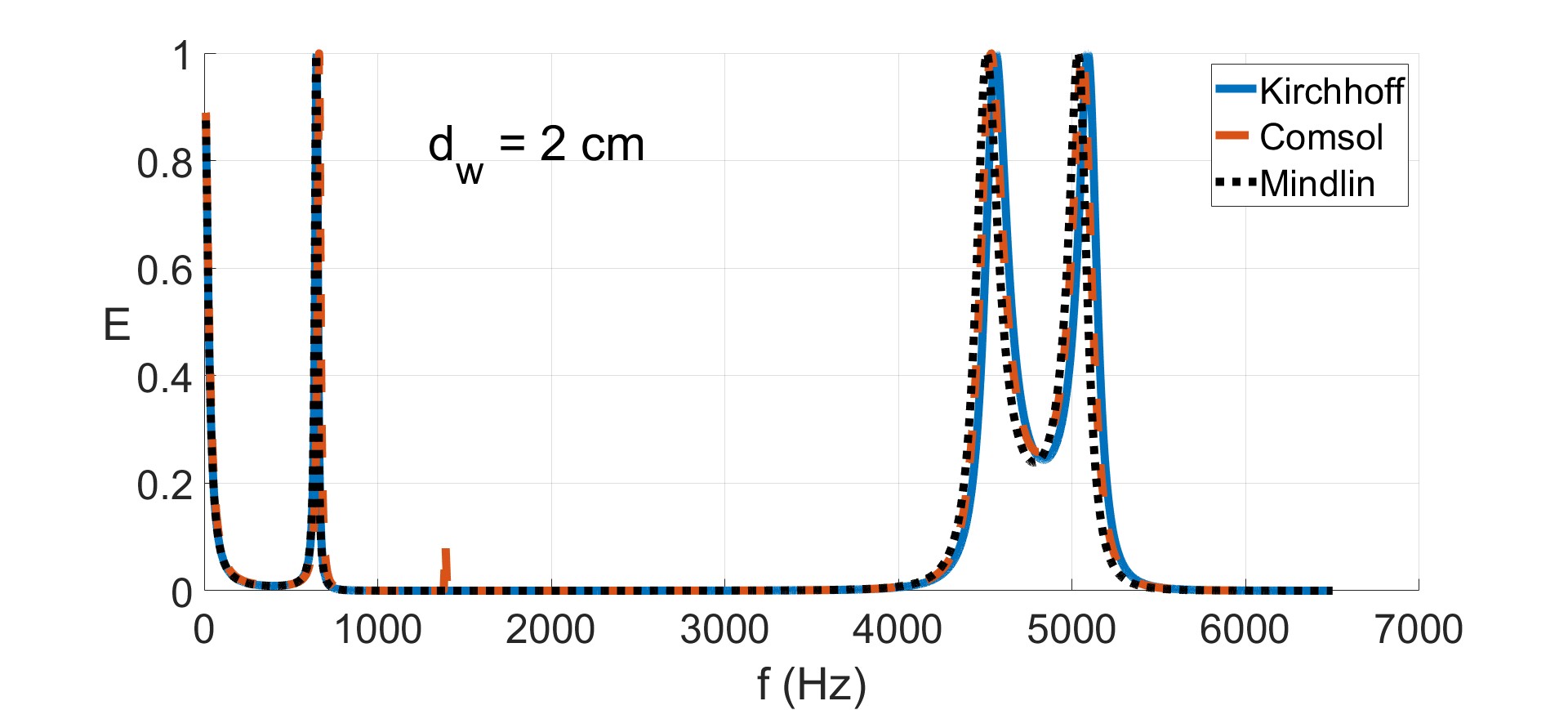}} 
    \subfigure[]{\includegraphics[width=0.49\textwidth]{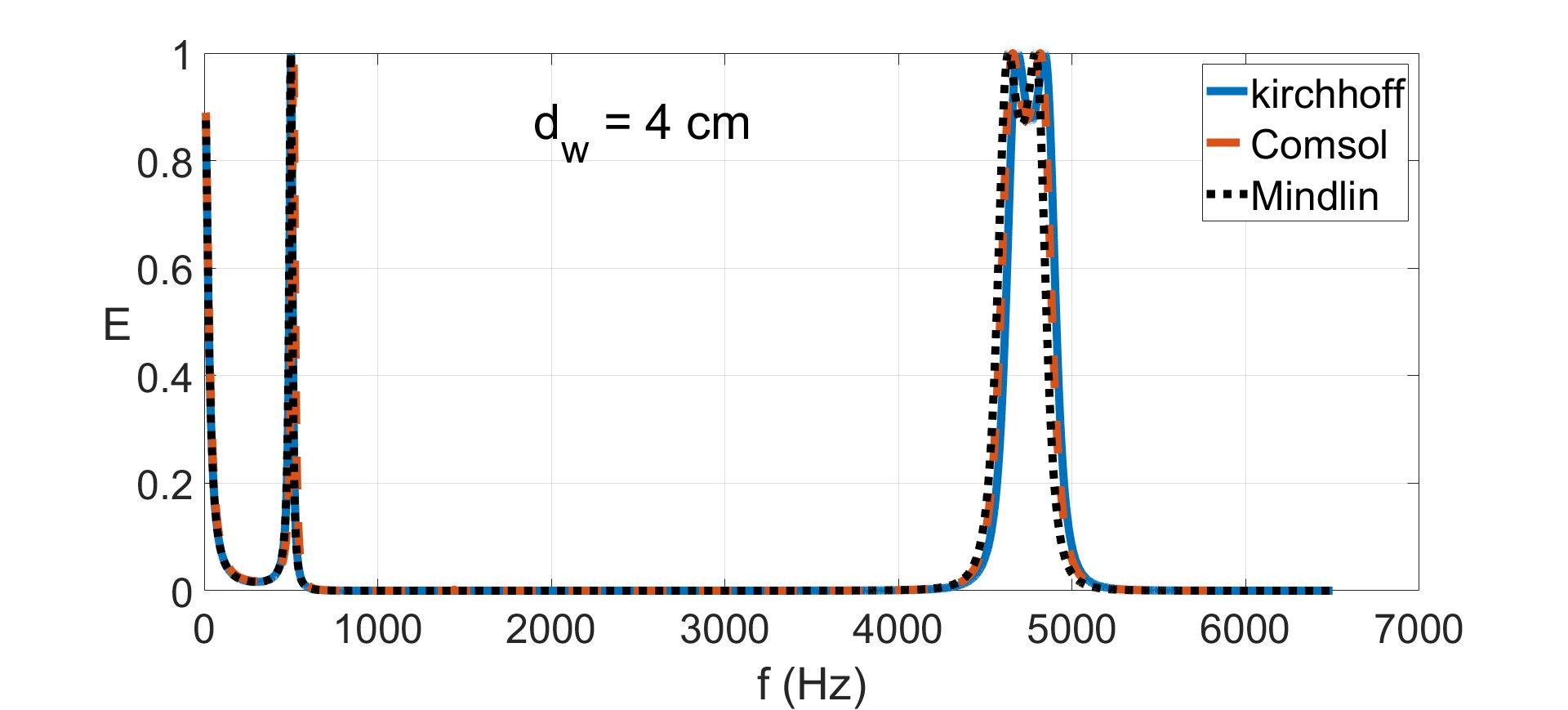}}
    \subfigure[]{\includegraphics[width=0.49\textwidth]{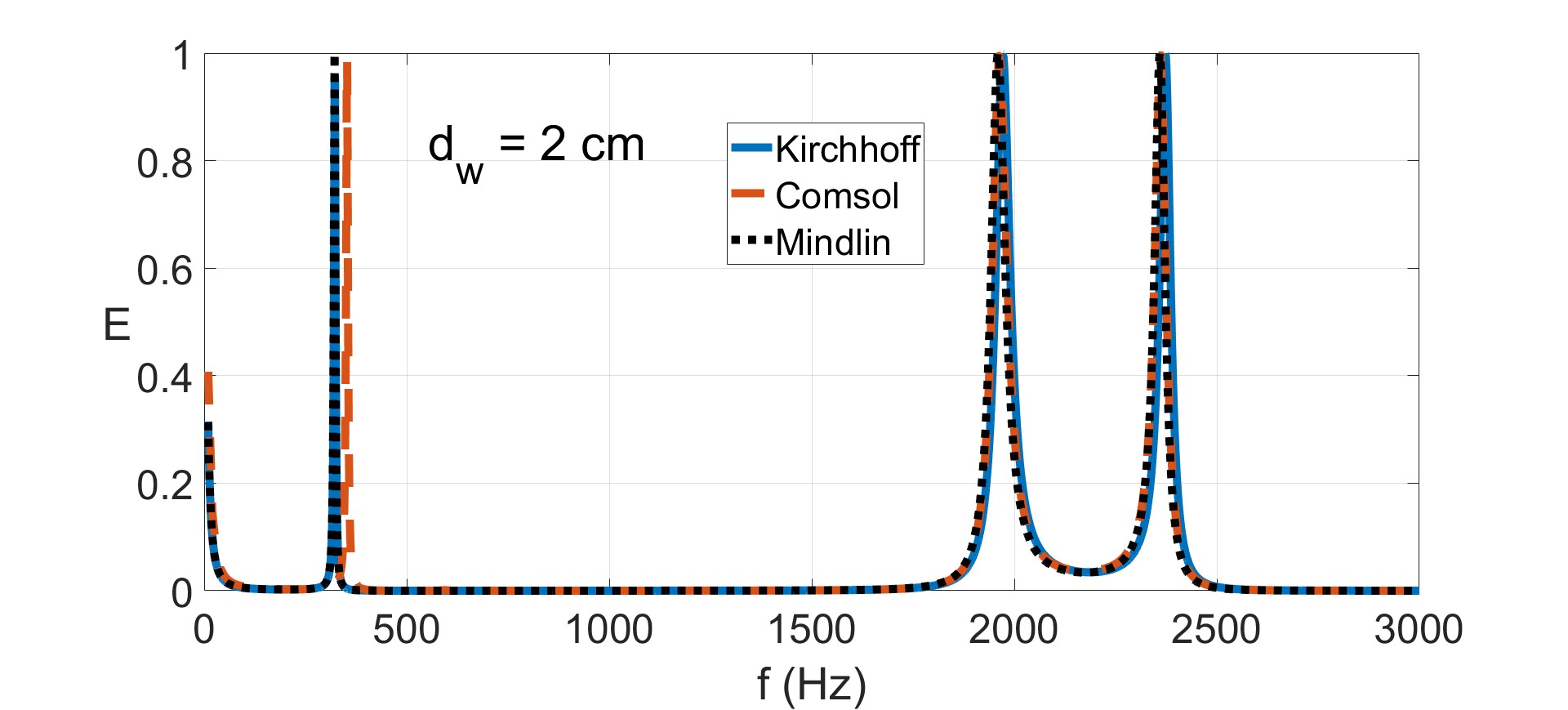}} 
    \subfigure[]{\includegraphics[width=0.49\textwidth]{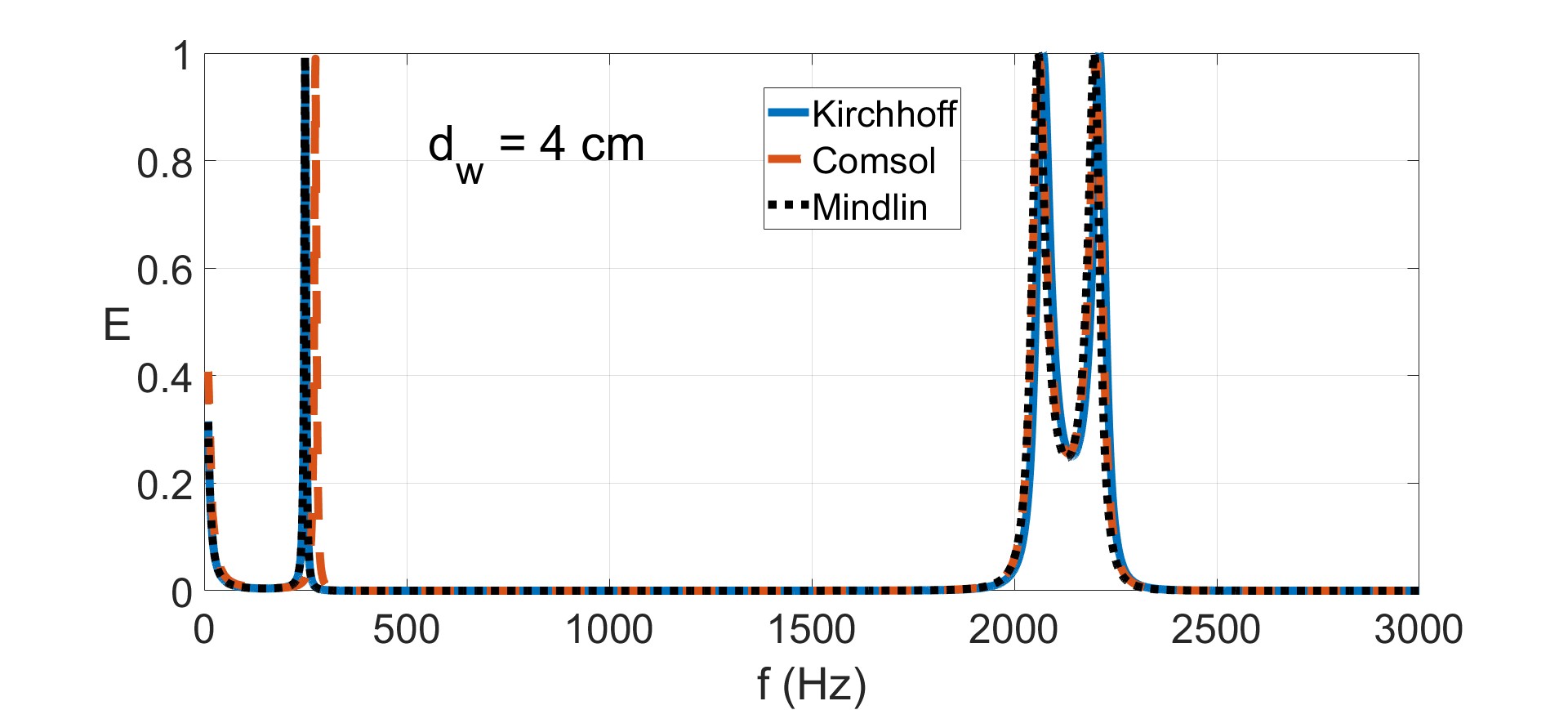}}
    \caption{The transmitted acoustic energy E vs frequency for the 2-Flex-layer model with a distance of $d_w$ from each other. The Aluminum plate thickness is considered 1 mm. (a,b) $b$ = 2.72 cm, (c,d) $b$ = 3.85 cm}
    \label{n2E}
\end{figure}
  \end{minipage} \end{widetext}                     

Figure (\ref{n2E}) shows the total transmitted energy (normalized to unity for incident).  The theory based on Kirchhoff and Mindlin models are compared with Comsol simulations.   The theoretical transmission coefficient $E$ is the propagating component of ${\bf u}^+$, that is 
\beq{-55}
E = T_{00} =  { {\bf u}^+_0 }^T    {\bf u}^+ .
\eeq
Comparing Figs.\ (\ref{n2E}) and  (\ref{n1E}) for $E$ reveals noticeable differences. It is observed that the first mode appears at lower frequencies in Fig.\ (\ref{n2E}). This observation can be explained by considering that the water between two flex structures has an equivalent mass, and the flex structures themselves possess equivalent mass and stiffness. Consequently, the first resonance of the 2-flex-layer model is formulated as $ f_1 \approx \frac{1}{2\pi} \sqrt{ \frac{2k_f}{2m_f + M_w} }$.  In addition to the low frequency transmission, Fig.\ \ref{n2E} also shows 
that the transmission modes of the single flex-layer, see Fig.\ \ref{n1E}, bifurcate into double transmission peaks.  The separation of the double peaks depends upon the spacing  $d_w$ between the flex-layers, with larger separation for smaller $d_w$.   This phenomenon is related to the presence of ``evanescent pressure" between the two flexible layers, which creates diffused angles of reflection on each flex-layer.    The bifurcation can also be interpreted as a "level splitting" effect due to the extra degree of freedom in the $n=2$ system.  

\begin{figure}[H]
    \centering
    \includegraphics[width=0.35\textwidth]{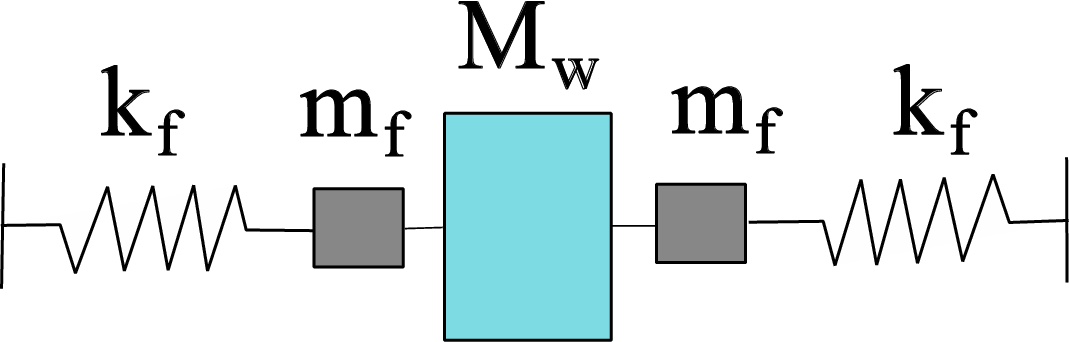}
    \caption{Simplified 2-flex-layer model for the mode $m = 1$}
    \label{W2FlexW_simplif}
\end{figure}

\section{Summary and conclusions} \label{sec6} 

The spatially periodic nature of the flex-layer causes it to generate an infinite set of evanescent  waves under plane wave incidence.  By solving for generalized Bragg wave incidence we have derived an analytical method to consider scattering from multiple flex-layers in series.  Although we have here restricted attention to the $n=2$ case it is clear that the method can be developed for an arbitrary  number of flex-layers by analogy with the purely one-dimensional case.  
The  solution is based on the explicit infinite-dimensional reflection and transmission matrices of Eq.\ \eqref{7=28}.   The scalar elements $R_{00}$ and $T_{00}$ provide the amplitudes of the far-field propagating reflected and transmitted waves.  The other elements define the evanescent contributions to the acoustic near-field.  

We have shown that the low frequency response of the flex-layer in water is analogous to that of a simple spring, which can be interpreted specifically in terms of an equivalent air gap.  Conversely, the flex-layer could be used to replace an air gap of given width in the sense of lumped acoustic elements.  The flex-layer exhibits total transmission (zero reflection) at finite frequencies defined by relation \eqref{3+5}.  While this must be solved numerically to find transmission frequencies, we have identified  the lowest one   as $f \approx 0.79\, c_f/b$ where $2b$ is the rib spacing and $c_f$ is the phase speed of the dispersive fluid-loaded flexural wave,  see Figs.\ \ref{bvsf} and \ref{cfvsf}.  

Finally, we have demonstrated the acoustic properties of  two   flex-layers in series, illustrating new effects not seen in the single flex-layer.  The mode splitting exhibited in Fig.\  \ref{n2E} can be controlled by the choice of the water separation between the flex-layers, leading to wider  total transmission peaks. This suggests that interesting broadband effects can be expected with multiple flex-layers, such as broadband total transmission with the potential for new applications in acoustic metamaterials.  Future work will examine multiple flex-layer systems, based upon the  reflection and transmission for the single layer introduced in this paper.




\appendix        

\section{Quasi-static  solution using Timoshenko theory}  \label{appa}
Instead of the Euler-Bernoulli theory \eqref{DWp0} we use the Timoshenko theory 
\beq{11}
D \phi '''(y) = p_0,
\ \ \
w'(y)  = \phi - \frac D{\mu h \kappa} \phi '' , 
\eeq
for $ -b \le y \le b$ 
where $\mu$ is the shear modulus and $\kappa$ is a shear correction factor. 
Solving  for the rotation  $\phi(y)$ and   plate deflection $w(y) $  
and setting $w(\pm b)=0$ and  $\phi  (\pm b)=0$,   gives 
$w(y) = w_\text{EB}(y)  +  p_0\, \big(b^2 - y^2 \big)/ (2\mu h \kappa )$
where $ w_\text{EB}$ is the deflection found in Section \ref{sec2} using Euler-Bernoulli beam theory.  
The magnitude of the deflection is therefore larger than that of the Euler-Bernoulli solution 
and yields 
\beq{4+21}
\Delta L = \frac{2}{45} \frac{b^4}{D}p_0  
\, \bigg( 1+   \frac {5h^2}{2b^2  \kappa (1-\nu)} \bigg)
\eeq
where we have used $E_p = 2\mu (1+\nu)$.  This $ \Delta L $  is    slightly greater as compared to that of Eq.\ \eqref{DVofFlex}, although  the relative correction is on the order of $10^{-3}$ for the values considered here.   In conclusion,  Timoshenko theory  gives very little difference for the frequencies considered. 

\section{Mindlin plate model} \label{appb} 

Using the Mindlin plate theory \cite{Stepanishen1978} changes $\hat Z_p$ from \eqref{7=9} to 
\beq{7=91}
\hat Z_p(k_y) = \frac { \big( Dk_y^2-\lambda \rho_s I \omega^2\big) \big(k_y^2-\frac{ \rho_s\omega^2}{\kappa \mu}\big)- \rho_s h \omega^2 }
{ -\ii \omega  \, \Big( 1  + \frac{D k_y^2-\lambda\rho_s I\omega^2}{\kappa \mu h} \Big)}
\eeq
where $\mu $ $(=\frac {E_p}{2(1+\nu)})$ is the shear modulus, $\kappa$ is a shear correction factor and $\lambda$ is rotary inertia correction factor.  Following \cite{Norris2018a}
we take  $\lambda = \kappa/\kappa_0$ and 
$\kappa = \frac{20}{17-7\nu} / \Big( 1 + 
\sqrt{1 - \frac {200 (1-\nu)}{\kappa_0 ({17-7\nu)^2} } }\Big)$
where $\kappa_0 = \pi^2 /12$.



\end{document}